\documentclass[aps,prd,onecolumn,groupedaddress,showpacs,nofootinbib,amssymb]{revtex4}
\usepackage[dvips]{graphicx}
\usepackage{amssymb}
\usepackage{amsmath}
\usepackage{graphicx,,color}
\usepackage{amsfonts}
\usepackage{bm}
\usepackage{cancel}
\usepackage{comment}
\newcommand\be{\begin{equation}}
\newcommand\ee{\end{equation}}
\newcommand\nn{\nonumber \\}
\newcommand\e{\mathrm{e}}

\allowdisplaybreaks[4]

\begin{document}

\title{The Reconstruction of $f(\phi)R$ and Mimetic Gravity from Viable Slow-roll Inflation}
\author{S.D. Odintsov,$^{1,2,3}$\,\thanks{odintsov@ieec.uab.es}
V.K. Oikonomou,$^{4,5}$\,\thanks{v.k.oikonomou1979@gmail.com}}
\affiliation{$^{1)}$ ICREA, Passeig Luis Companys, 23, 08010 Barcelona, Spain\\
$^{2)}$ Institute of Space Sciences (IEEC-CSIC) C. Can Magrans s/n,
08193 Barcelona, Spain\\
$^{3)}$ Tomsk State Pedagogical University, 634061 Tomsk, Russia\\
$^{4)}$Department of Physics, Aristotle University of Thessaloniki, Thessaloniki 54124, Greece\\
$^{5)}$ Laboratory for Theoretical Cosmology, Tomsk State University
of Control Systems
and Radioelectronics, 634050 Tomsk, Russia (TUSUR)\\
}

\tolerance=5000

\begin{abstract}
In this work, we extend the bottom-up reconstruction framework of $F(R)$ gravity to other modified gravities, and in particular for $f(\phi)R$ and mimetic $F(R)$ gravities. We investigate which are the important conditions in order for the method to work, and we study several viable cosmological evolutions, focusing on the inflationary era. Particularly, for the $f(\phi)R$ theory case, we specify the functional form of the Hubble rate and of the scalar-to-tensor ratio as a function of the $e$-foldings number and accordingly, the rest of the physical quantities and also the slow-roll and the corresponding observational indices can be calculated. The same method is applied in the mimetic $F(R)$ gravity case, and in both cases we thoroughly analyze the resulting free parameter space, in order to show that the viability of the models presented is guaranteed and secondly that there is a wide range of values of the free parameters for which the viability of the models occurs. In addition, the reconstruction method is also studied in the context of mimetic $F(R)=R$ gravity. As we demonstrate, the resulting theory is viable, and also in this case, only the scalar-to-tensor ratio needs to be specified, since the rest follow from this condition. Finally, we discuss in brief how the reconstruction method could function for other modified gravities.
\end{abstract}

%PACS numbers: 04.50.Kd, 95.36.+x, 98.80.-k, 98.80.Cq

\maketitle

\section{Introduction}

The last thirty years the observations coming from the Universe have offered many exciting moments to the scientific community. The first astonishing event was the observation of the late-time acceleration in the late 90's \cite{Riess:1998cb}, and thereafter, the Planck observations on the early-time evolution \cite{Ade:2015lrj}, the observation of the merging of black holes which created gravitational waves \cite{Connaughton:2016umz}, and lately the observation of the two neutron stars merging \cite{TheLIGOScientific:2017qsa}. The latter was particularly exciting, since we received both gravitational and electromagnetic waves. This offers fruitful phenomenological data, and with these the scientific community may impose stringent bounds on various theoretical frameworks \cite{Nojiri:2017hai}. In close connection to these issues, the lower mass limit of neutron stars may be even further constrained \cite{Bauswein:2017vtn}, which in effect may favor some modified gravity theories which predict massive neutron stars \cite{Capozziello:2015yza,Astashenok:2014nua} which cannot be described by the Einstein-Hilbert framework.

In this line of research, the primordial era and the events generated during it, is one of the primary interests of cosmologists. The current constraints of Planck on the primordial era, and also the possible future observations, may shed light on the most mysterious era of our Universe's existence. One appealing candidate theory that describes in an accurate way the early-time era of our Universe is the theory of inflation \cite{inflation1,inflation2,inflation3,inflation4,reviews1}, which remedies many flaws of the standard Big-Bang cosmology. There are many theoretical frameworks which can successfully describe inflation, such as scalar-tensor approaches \cite{inflation1} and also modified gravity approaches \cite{reviews1,reviews2,reviews4,reviews5,reviews6,reviews7}, for which in many cases the compatibility with the Planck data \cite{Ade:2015lrj} may be achieved. The other alternative theory that may also describe the primordial phases of our Universe is the bounce cosmology, which may also describe successfully in some cases the evolution of the Universe, see Refs.  \cite{Brandenberger:2016vhg,Cai:2014bea,Cai:2015emx,Cai:2016hea,Lehners:2011kr} for reviews and also \cite{Cai:2012va,Cai:2011tc,Cai:2013kja,Cai:2014jla,Lehners:2015mra,Koehn:2015vvy,Odintsov:2015zza} for an important stream of papers. Moreover, combination of inflationary and bounce cosmologies may also yield appealing results \cite{Liu:2013kea,Piao:2003zm}.

Due to the importance of the inflationary theories, it is compelling to have a considerable command on the theoretical frameworks that may successfully describe inflation, and also it is even more important to have a general guide on how to generate viable theories of inflation. To this end, in this paper we shall be interested in providing a general reconstruction method for some classes of modified gravity \cite{reviews1,reviews2,reviews4,reviews5,reviews6,reviews7} which may generate viable inflationary evolutions. Our approach will be a bottom-up method, in which starting from the scalar-to-tensor ratio, we shall fix it to have a desirable form, and from it we shall investigate which modified gravity can generate such a cosmology. In a recent work, we presented how this reconstruction technique may yield viable $F(R)$ gravities, so in this work we generalize the bottom-up method to include alternative modified gravities. An alternative reconstruction scheme to ours, was recently presented by A. Starobinsky in \cite{starobconf}. Also there are also other alterative approaches to ours in the literature, that in some cases use the renormalization group approach \cite{Narain:2017mtu,Narain:2009fy,Fumagalli:2016sof} for $F(R)$ gravity, see also \cite{Gao:2017uja,Lin:2015fqa,Miranda:2017juz,Fei:2017fub}, for scalar-tensor reconstruction  from the observational indices theories.

In this work we shall be interested in $f(\phi)R$ and mimetic $F(R)$ gravity theories. The attributes of $f(\phi)R$ theories are well-known, and also the mimetic framework \cite{Chamseddine:2013kea} has recently become particularly well known, due to the  various astrophysical and cosmological applications that this framework has. The mimetic framework was further studied in Refs. \cite{Chamseddine:2014vna,Hammer:2015pcx,Golovnev:2013jxa,Matsumoto:2015wja,Leon:2014yua,Haghani:2015zga,Cognola:2016gjy,Nojiri:2014zqa,Raza:2015kha,Odintsov:2015wwp,Odintsov:2015cwa,Chen:2017ify,Shen:2017rya,Gorji:2017cai,Bouhmadi-Lopez:2017lbx,Takahashi:2017pje,Vagnozzi:2017ilo,Baffou:2017pao,Sebastiani:2016ras}
and also for a recent concise review on mimetic gravity see \cite{Sebastiani:2016ras}. To our opinion, the most appealing features of mimetic gravity \cite{Chamseddine:2013kea}, and also of mimetic $F(R)$ gravity \cite{Nojiri:2014zqa}, is firstly the fact that dark matter may be described in a geometric way without the need of the particle description \cite{Oikonomou:2006mh}, and secondly, in many cases a viable inflationary era may be predicted, which can be in concordance with the latest Planck \cite{Ade:2015lrj} and also with the BICEP2/Keck-Array data \cite{Array:2015xqh}. Since the future observational data can even further constrain the available theories, it is compelling to have alternative modified gravities which may pass the current tests, but also leave room for even further stringent constraints. As we demonstrate, it is possible to have such theories and in this work we shall provide a general method on how to obtain such theories. Particularly, we review in brief how the bottom-up reconstruction method works in the case of $F(R)$ gravity and we present the functional form of the potential of the scalar-tensor counterpart theory. After this we study the $f(\phi)R$ theory and we present the bottom-up reconstruction method in this case. For the $f(\phi)R$ case, in order for the reconstruction method to work, it is required that the Hubble evolution is given, after choosing this appropriately in order for the accelerating phase to occur. So by fixing the Hubble rate and the scalar-to-tensor ratio, we obtain a differential equation which specifies the form of $f(\phi(N))$. Accordingly, we find the functions $V(N)$ and $\phi(N)$, and by inverting the latter, we can find the analytic forms of $V(\phi)$ and $f(\phi)$. After having all the above quantities available, we demonstrate that the resulting inflationary evolution is viable, in the context of the slow-roll approximation, by calculating the slow-roll indices and the corresponding observational indices, focusing on the spectral index of the primordial curvature perturbations and the already specified scalar-to-tensor ratio. Also we demonstrate that the viability of the model can be achieved for a wide range of the free parameters values. After the $f(\phi)R$ theory, we discuss the mimetic $F(R)$ gravity, and we follow the same procedure, by specifying the Hubble rate $H(N)$ and the scalar-to-tensor ratio as a function of the $e$-foldings. After this we solve the resulting differential equations and we specify the $F(R(N))$ gravity. Also by using the analytic form of the scalar curvature, we can specify the function $F(R)$ and from it the function $f(R)$. Moreover, we specify the functional forms of the scalar potential and of the Lagrange multiplier corresponding to the mimetic theory. Having all the above quantities at hand, enables us to calculate the slow-roll indices and the corresponding spectral index of primordial curvature perturbations and the scalar-to-tensor ratio and we demonstrate that the resulting inflationary theory is viable. In addition, we show that in this case too, the viability can be achieved for a wide range of values of the free parameters. In all the aforementioned cases, special attention is given on the allowed parameter values, since the Hubble evolution must indeed describe an inflationary theory. Finally, in Appendix B, we address the $F(R)=R$ mimetic gravity case and we discuss the viability of the model.

This paper is organized as follows: In section II we review in brief how the bottom-up reconstruction method works in the case of $F(R)$ gravity, focusing mainly on presenting the main features of the method and the results. Also we demonstrate that a viable inflationary evolution is possible in this case. Also we present the Einstein frame scalar potential of the scalar counterpart theory. In section III we study the bottom-up reconstruction method in the case of $f(\phi)R$ gravity. We use some appropriately chosen cosmological evolutions and we discuss how the viability can be achieved in this case. Also the viability of this model is thoroughly investigated. In section IV we perform the same analysis in the case of mimetic $f(R)$ gravity, and we demonstrate that our bottom-up approach leads to a viable inflationary evolution. For all the models we study, we present detailed formulas of the various physical quantities and of the slow-roll indices, as functions of the $e$-foldings number. In this way, the viability of the models becomes quite easy to check in a straightforward manner. Finally, the conclusions along an informative discussion on the potential applications, appear in the end of the paper.

Before we start, let us briefly present the geometric background which we shall assume. Particularly, we assume that the background geometry has the following line element,
\begin{equation}
\label{JGRG14} ds^2 = - dt^2 + a(t)^2 \sum_{i=1,2,3}
\left(dx^i\right)^2\, ,
\end{equation}
which is a flat Friedmann-Robertson-Walker (FRW) geometric background, with $a(t)$ being the scale factor. Also, the metric connection is assumed to be a symmetric and torsion-less metric-compatible connection, the Levi-Civita connection.

\section{Reconstruction of Viable Inflationary $F(R)$ Gravity}

In a previous work of ours \cite{Odintsov:2017fnc}, we used the bottom up approach in order to find which $f(R)$ gravity may produce a specific scalar-to-tensor ratio. In this paper we aim to generalize this bottom up approach to other modified gravities. However, before we get into the core of this paper, for completeness it is worth presenting in brief the results of the $F(R)$ gravity case. For details we refer the read to Ref. \cite{Odintsov:2017fnc}. In the slow-roll approximation, the slow-roll indices $\epsilon_1$ ,$\epsilon_2$, $\epsilon_3$, $\epsilon_4$, for the pure $F(R)$ gravity theory read \cite{Noh:2001ia,Hwang:2001qk,Hwang:2001pu,Kaiser:2010yu,reviews1},
\begin{equation}
\label{restofparametersfr} \epsilon_1=-\frac{\dot{H}}{H^2},\,\,\,\epsilon_2=0\, ,\quad \epsilon_1\simeq
 -\epsilon_3\, ,\quad \epsilon_4\simeq
 \frac{F_{RRR}}{F_{R}}\left( 24\dot{H}+6\frac{\ddot{H}}{H}\right)-3\epsilon_1+\frac{\dot{\epsilon}_1}{H\epsilon_1}\,
 ,
\end{equation}
with $F_R=\frac{\mathrm{d}F}{\mathrm{d}R}$, and
$F_{RRR}=\frac{\mathrm{d}^3F}{\mathrm{d}R^3}$. Moreover, the observational indices $n_s$ and $r$ are \cite{Noh:2001ia,Hwang:2001qk,Hwang:2001pu,Kaiser:2010yu,reviews1},
\begin{equation}
\label{epsilonall} n_s\simeq 1-6\epsilon_1-2\epsilon_4,\quad
r\simeq 48\epsilon_1^2\, ,
\end{equation}
where $n_s$ is the spectral index of the primordial curvature perturbations and $r$ is the scalar-to-tensor ratio. The fact that the scalar-to-tensor ratio $r$ is equal to the square of the first slow-roll index $\epsilon_1$ may seem a bit strange, since in most cases of single scalar field inflation theories, $r$ is proportional to $\epsilon_1$. This is the source of the difference, since we are considering a Jordan frame $f(R)$ gravity, so the result applies only for $f(R)$ gravity and also when the slow-roll approximation is used. It is worth proving this vague issue, although it appears in standard texts in the field, see \cite{reviews1}, so in the Appendix A, we provide a proof for the expression of $r$ in the case of $f(R)$ gravity.

As it was shown in Ref. \cite{Odintsov:2017fnc}, the bottom up reconstruction method is based on the assumption that the scalar-to-tensor ratio has an appropriately chosen functional form, so we adopt the choice of \cite{Odintsov:2017fnc}, and we set,
\begin{equation}\label{model1}
r=\frac{c^2}{(q+N)^2}\, ,
\end{equation}
where $N$ denotes the $e$-foldings number and $c$, $q$ are free parameters. The scalar-to-tensor ratio as a function of $N$ is,
\begin{equation}\label{diffsqueareN}
r=\frac{48 H'(N)^2}{H(N)^2}\, ,
\end{equation}
where the prime indicates differentiation with respect to the $e$-foldings number $N$.
By using Eqs. (\ref{model1}) and (\ref{diffsqueareN}), we get,
\begin{equation}\label{diffeqns}
\frac{\sqrt{48} H'(N)}{H(N)}=\frac{c}{(q+N)}\, ,
\end{equation}
which when solved it yields,
\begin{equation}\label{hubbleratesol}
H(N)=\gamma  (N+q)^{\frac{c}{4 \sqrt{3}}}\, .
\end{equation}
Also the spectral index as a function of $N$ is,
\begin{equation}\label{spectralnhnexpr}
n_s\simeq 1+\frac{4 H'(N)}{H(N)}-\frac{2 \left(H(N)
H''(N)+H'(N)^2\right)}{H(N) H'(N)}+ \frac{F_{RRR}}{F_{R}}\Big{(}
24H(N)H'(N)+6H(N)H''(N)+6H'(N)^2\Big{)}\, ,
\end{equation}
so by using Eq.
(\ref{hubbleratesol}), we obtain,
\begin{align}\label{spctralindexfinal}
& n_s=1+\frac{c}{\sqrt{3} (N+q)}-\frac{c N}{\sqrt{3}
(N+q)^2}-\frac{c q}{\sqrt{3} (N+q)^2}+\frac{2 N}{(N+q)^2}+\frac{2
q}{(N+q)^2}+\\ \notag & \frac{c^2 \gamma ^2 F_{RRR}
(N+q)^{\frac{c}{2 \sqrt{3}}-2}}{8 F_R}+\frac{5 \sqrt{3} c \gamma ^2
F_{RRR}(N+q)^{\frac{c}{2 \sqrt{3}}-1}}{2 F_R}\, .
\end{align}
In Ref. \cite{Odintsov:2017fnc}, it was demonstrated that since the Hubble rate is given as a function of the $e$-foldings number $N$, the $F(R)$ gravity can be found by using well-known reconstruction techniques \cite{Nojiri:2009kx}, and accordingly the viability of the theory can be checked. The resulting $F(R)$ gravity which generates the evolution (\ref{hubbleratesol}) can be found in closed form only in the case that $c=\sqrt{12}$, and it is equal to,
\begin{equation}\label{frtermafinal}
F(R)=R-\frac{\gamma ^2}{2}-\frac{R^2}{18 \gamma ^2}\, .
\end{equation}
which is an alternative form of the well-known Starobinsky model
\cite{Starobinsky:1982ee}. Also, similar forms of $F(R)$ gravity were studied in \cite{Sebastiani:2013eqa}. Due to the fact that , the spectral index as a function of $N$, is equal to,
\begin{align}\label{spctralindexfinalertermsilan}
& n_s=1+\frac{c}{\sqrt{3} (N+q)}-\frac{c N}{\sqrt{3}
(N+q)^2}-\frac{c q}{\sqrt{3} (N+q)^2}+\frac{2 N}{(N+q)^2}+\frac{2
q}{(N+q)^2}\, ,
\end{align}
so by choosing $c=\sqrt{12}$, $N=60$
and $q=-118$, we obtain,
\begin{equation}\label{observations1plancbicepresults}
n_s\simeq 0.9658,\,\,\,r\simeq 0.00346842\, .
\end{equation}
The 2015 Planck data constrain the observational indices
in the following way,
\begin{equation}
\label{planckdata} n_s=0.9644\pm 0.0049\, , \quad r<0.10\, ,
\end{equation}
and in addition, the latest BICEP2/Keck-Array data \cite{Array:2015xqh}
indicate that,
\begin{equation}
\label{scalartotensorbicep2} r<0.07\, ,
\end{equation}
at $95\%$ confidence level. Hence the resulting values of the observational indices (\ref{observations1plancbicepresults}), are in concordance with both the Planck and the BICEP2/Keck-Array data.

Finally we need to show that the cosmological evolution (\ref{hubbleratesol}) satisfies $\ddot{a}>0$, and hence it is an accelerating cosmology, where the double ``dots'' indicate differentiation with respect to the cosmic time. Thus we need the Hubble rate expressed in terms of the cosmic time, so we need to solve the following differential equation,
\begin{equation}\label{nthubbleast}
\dot{N}=H(N(t))\, ,
\end{equation}
and by using (\ref{hubbleratesol}) we obtain,
\begin{equation}\label{hubblecaseonecosmict}
N(t)=\frac{1}{4} \left(\Lambda ^2-4 q+\gamma ^2 t^2-2 \gamma \Lambda
t\right)\, ,
\end{equation}
with $\Lambda>0$ being a free parameter. Then Eq. (\ref{hubblecaseonecosmict}) in conjunction with Eq. (\ref{hubbleratesol}) , yield,
\begin{equation}\label{cosmictimes}
H(t)=\frac{\gamma  \Lambda }{2}-\frac{\gamma ^2 t}{2}\, .
\end{equation}
Therefore, it is easy to show that the resulting evolution yields $\ddot{a}>0$.

It is easy to obtain the scalar-tensor counterpart theory from the $F(R)$ gravity theory, by using a conformal transformation which will bring the Jordan frame quantities to their Einstein frame counterpart. The canonical transformation that connects the Jordan with the Einstein frame is,
\begin{equation}
\label{can}
\varphi =\sqrt{\frac{3}{2}}\ln (F'(R))\, ,
\end{equation}
so by applying this transformation in the Jordan frame action of $F(R)$ gravity, one obtains the Einstein frame potential, which is,
\begin{align}
\label{potentialvsigma}
V(\phi)
=\frac{1}{2}\left(\frac{R}{F'(R)}-\frac{F(R)}{F'(R)^2}\right)=\frac{1}{2}\left
( \e^{-\sqrt{2/3}\phi }R\left (\e^{\sqrt{2/3}\phi} \right )
 - \e^{-2\sqrt{2/3}\phi }F\left [ R\left (\e^{\sqrt{2/3}\phi}
\right ) \right ]\right )\, .
\end{align}
Let us investigate how the Einstein frame potential looks like for the $F(R)$ gravity of Eq. (\ref{frtermafinal}), so by substituting the functional form of the $F(R)$ gravity in Eq. (\ref{potentialvsigma}), we obtain, the following scalar potential,
\begin{equation}\label{scalarpotentialeinsteinfr}
V(\phi)=\frac{9}{2} \gamma ^2 e^{-\sqrt{\frac{2}{3}} \phi }-\frac{5}{2} \gamma ^2 e^{-2 \sqrt{\frac{2}{3}} \phi }-\frac{9 \gamma ^2}{4}\, .
\end{equation}
In principle, the observational indices for the Einstein frame scalar theory and for the Jordan frame $F(R)$ theory should coincide \cite{Kaiser:1995nv,Brooker:2016oqa}, however the calculations in the Einstein frame must be performed carefully at leading order since the form of the potential makes compelling to us an approximate form for the potential. Nevertheless, the conformal invariant quantities, for example the spectral index $n_s$, and the scalar-to-tensor ratio $r$ are the same in both the $F(R)$ frame and the corresponding Einstein frame.

\section{Reconstruction of Viable Inflationary $f(\phi)R$ Gravity}

In this section we consider $f(\phi)R$
theories of gravity, which describe non-minimally coupled theories
of gravity, in which case the action is,
\begin{equation}
\label{graviactionnonminimal}
\mathcal{S}=\int d ^4x\sqrt{-g}\left(
\frac{f(\phi)R}{2\kappa^2}-\frac{1}{2}g^{\mu\nu}
\partial_{\mu}\phi\partial_{\nu}\phi-V(\phi) \right)\, ,
\end{equation}
where $f(\phi)$ is an analytic function of the scalar field $\phi$. In the literature there are various works for this class of modified gravity models, see \cite{Comer:1996pq,Myrzakul:2015gya,Myrzakulov:2015qaa}. By varying the action (\ref{graviactionnonminimal}) with respect to
the metric and with respect to the scalar field $\phi$, we obtain
the following equations of motion,
\begin{equation}
\label{gravieqnsnonminimal}
\frac{3f}{\kappa^2}H^2=\frac{\dot{\phi}^2}{2}
+V(\phi)-3h\frac{\dot{f}}{\kappa^2}\, ,\quad
 -\frac{2f}{\kappa^2}\dot{H}=\dot{\phi}^2
+\frac{\ddot{f}}{\kappa^2}-H\frac{\dot{f}}{\kappa^2}\, ,\quad
\ddot{\phi}+3H\dot{\phi}
 -\frac{1}{2\kappa^2}R\frac{df}{d \phi}+\frac{d V}{d \phi}=0\, ,
\end{equation}
with the ``dot'' denoting as usual, differentiation with respect
to the cosmic time. The slow-roll indices in the case of a
non-minimally coupled scalar theory, are equal to,
\begin{equation}
\label{slowrollnonminimal}
\epsilon_1=-\frac{\dot{H}}{H^2}\, ,\quad
\epsilon_2=\frac{\ddot{\phi}}{H\dot{\phi}}\, ,\quad
\epsilon_3=\frac{\dot{f}}{2Hf} \, ,\quad
\epsilon_4=\frac{\dot{E}}{2HE}\, ,
\end{equation}
and in this case the function
$E$ is equal to,
\begin{equation}
\label{epsilonparameter}
E=f+\frac{3\dot{f}^2}{2\kappa^2\dot{\phi}^2}\, .
\end{equation}
The spectral
index of primordial curvature perturbations $n_s$ and the
scalar-to-tensor ratio in terms of the slow-roll indices, are equal
to,
\begin{equation}
\label{observatinalindices1}
n_s\simeq
1-4\epsilon_1-2\epsilon_2+2\epsilon_3-2\epsilon_4\, ,\quad
r=8\kappa^2\frac{Q_s}{f}\, ,
\end{equation}
and it is assumed that the
slow-roll indices satisfy the slow-roll condition $\epsilon_i\ll 1$,
$i=1,..,4$. In addition, the parameter $Q_s$ in the case at hand is
equal to,
\begin{equation}
\label{qs1nonslowroll}
Q_s=\dot{\phi}^2\frac{E}{fH^2(1+\epsilon_3)^2}\, .
\end{equation}
We can find
an approximate expression for the function $Q_s$, by using the
slow-roll approximation, so we find the slow-roll approximated
gravitational equations of motion, which take the following form,
\begin{align}
\label{gravieqnsslowrollapprx12}
& \frac{3fH^2}{\kappa^2}\simeq V(\phi)\, ,\quad
%\label{gravieqnsslowrollapprx2}
%&
3H\dot{\phi}-\frac{6H^2}{\kappa^2}f'+V'\simeq 0\, , \\
\label{gravieqnsslowrollapprx3}
& \dot{\phi}^2\simeq
\frac{H\dot{f}}{\kappa^2}-\frac{2f\dot{H}}{\kappa^2}\, ,
\end{align}
with the ``prime'' denoting this time differentiation with respect
to the scalar $\phi$. In view of the slow-roll approximated
equations of motion, the parameter $Q_s$ becomes,
\begin{equation}
\label{qs2}
Q_s=\frac{\dot{\phi}^2}{H^2}+\frac{3\dot{f}^2}{2\kappa^2fH^2}\, ,
\end{equation}
and in conjunction with Eq.~(\ref{gravieqnsslowrollapprx3}), the
parameter $Q_s$ finally becomes,
\begin{equation}
\label{qs3}
Q_s\simeq
\frac{H\dot{f}}{H^2\kappa^2}-\frac{2f\dot{H}}{\kappa^2H^2}\, .
\end{equation}
Hence, by combining Eqs.~(\ref{observatinalindices1}) and
(\ref{qs3}), we can find a simplified expression for the
scalar-to-tensor ratio during the slow-roll era, which is,
\begin{equation}
\label{rscalartensornoniminal}
r\simeq 16 (\epsilon_1+\epsilon_3)\, .
\end{equation}
Moreover, we may express the spectral index of the primordial
curvature perturbations $n_s$ as a function of the slow-roll indices
in the slow-roll approximation, which takes the following form,
\begin{equation}
\label{nsintersmofslowroll}
n_s\simeq 1-2\epsilon_1\left (\frac{3H\dot{f}}{\dot{\phi}^2}+2
\right)-2\epsilon_2-6\epsilon_3\left(
\frac{H\dot{f}}{\dot{\phi}^2}-1\right)\, .
\end{equation}
The formalism we just developed, enables us to calculate
analytically the observational indices during the slow-roll era, for
any given function $f(\phi)$.

In the following we use a physical units system where $\kappa=1$. Let us investigate how the slow-roll indices and the corresponding observational indices become when these are expressed in terms of the $e$-foldings number $N$. The slow-roll indices $\epsilon_1(N)$, $\epsilon_2(N)$ and $\epsilon_3(N)$, in terms of $N$ become,
\begin{align}\label{slowrollindicesfrphin}
& \epsilon_1(N)=-\frac{H'(N)}{H(N)},\,\,\, \epsilon_2(N)=\frac{H'(N)^2+H(N)^2 \phi ''(N)}{H(N)^2 \phi '(N)},\,\,\,\epsilon_3(N)=\frac{f'(N)}{2 f(N)}\, ,
\end{align}
so the spectral index of the primordial curvature perturbations becomes,
\begin{equation}\label{spcectrralindexnfrphi}
n_s=-2 \epsilon_1(N) \left(\frac{3 H(N)^2 f'(N)}{H(N)^2 \phi '(N)^2}+2\right)-6 \epsilon_3(N) \left(\frac{H(N)^2 f'(N)}{H(N)^2 \phi '(N)^2}-1\right)-2 \epsilon_2(N)+1\, ,
\end{equation}
and the scalar-to-tensor ratio is simply,
\begin{equation}\label{scalartotensornfrphi}
r=16 (\epsilon_1(N)+\epsilon_3(N))\, .
\end{equation}
Now let us explain in detail the reconstruction technique in the case of $f(\phi)R$ gravity and it goes as follows: First find a suitably chosen cosmological evolution $H(N)$ that it's functional form and parameters will be chosen in order the scale factor with respect to the cosmic time satisfies eventually $\ddot{a}>0$, which is the condition for an inflationary evolution. The second step of the reconstruction technique is to assume that the scalar-to-tensor ratio has a specific desirable form. In the $f(\phi)R$ gravity case, by assuming that the scalar-to-tensor ratio is equal to a function $g(N)$, that is $r=g(N)$, appropriately chosen, by using Eqs. (\ref{slowrollindicesfrphin}) and (\ref{scalartotensornfrphi}), we obtain the differential equation,
\begin{equation}\label{diffeqtionfrphiscalar}
16 \left(\frac{f'(N)}{2 f(N)}-\frac{H'(N)}{H(N)}\right)=g(N)\, .
\end{equation}
Then by solving the above differential equation, one obtains the function $f(N)$, in an analytic form. Then the function $\phi (N)$ may be obtained, by solving the slow-roll differential equation (\ref{gravieqnsslowrollapprx3}), which in terms of the $e$-foldings number $N$ it can be expressed as follows,
\begin{equation}\label{slowrollequatidifffrphi}
\left(H(N) \phi '(N)\right)^2=H(N)^2 f'(N)-2 f(N) H(N)H'(N)\, .
\end{equation}
Once the function $\phi (N)$ is obtained, the slow-roll index $\epsilon_2(N)$ can be obtained, and from it, the spectral index of the primordial curvature perturbations $n_s$ appearing in Eq. (\ref{spcectrralindexnfrphi}), can be calculated. At this step, the viability of the theory can be investigated explicitly, by appropriately adjusting the parameters of the theory. Eventually, the scalar potential of the theory can also be obtained in the slow-roll approximation, by inverting the function $\phi (N)$, and by using the first relation in Eq. (\ref{gravieqnsslowrollapprx12}).

\begin{figure}[h]
\centering
\includegraphics[width=18pc]{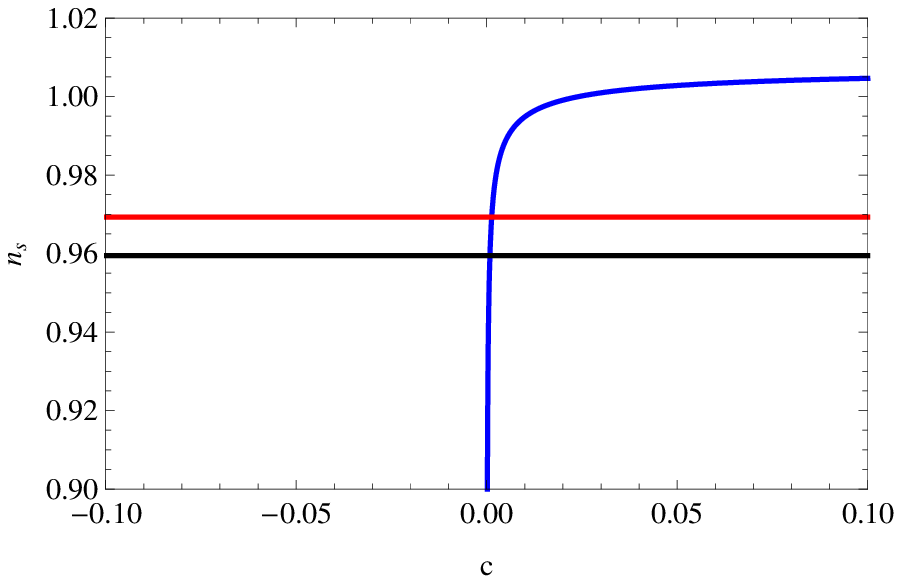}
\includegraphics[width=18pc]{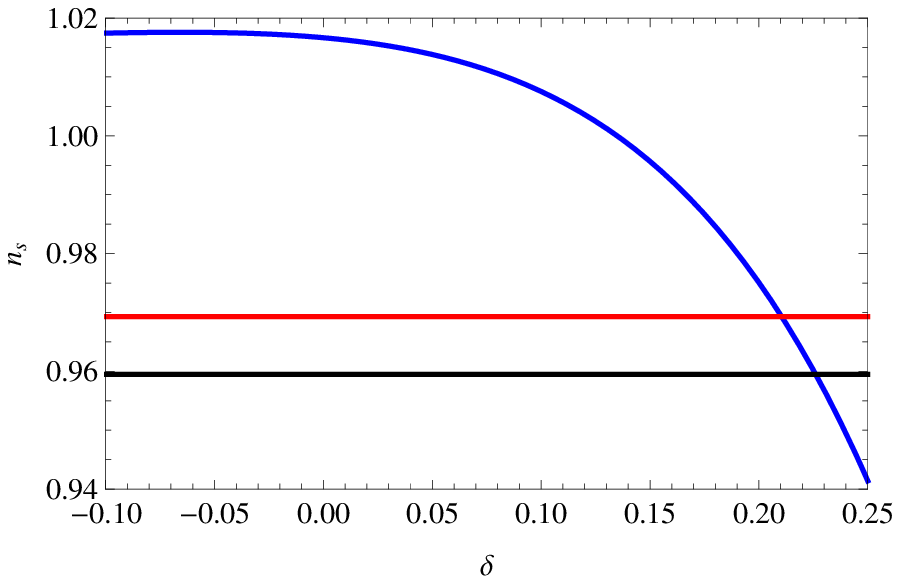}
\includegraphics[width=18pc]{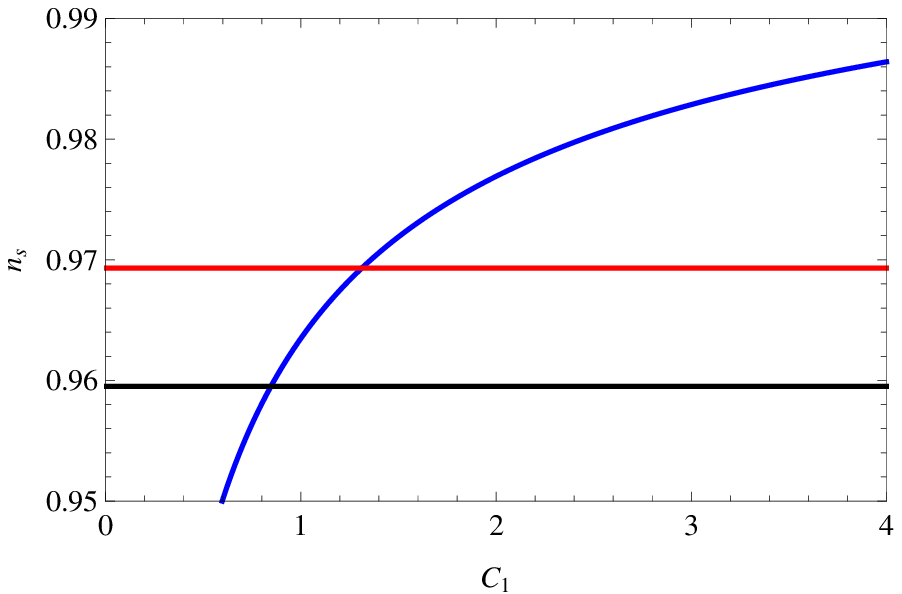}
\caption{The behavior of the spectral index $n_s$ as a function of the parameter $c$ (upper left plot), for $N=60$, $\delta=0.22$, $\gamma=30$, $C_1=1$, as a function of $\delta$ (upper right plot) for $N=60$, $c=0.001$, $\gamma=30$, $C_1=1$, and as a function of $C_1$ (bottom plot) for $N=60$, $\delta=0.22$, $\gamma=30$, $c=0.001$. In all the plots, the dependence of the spectral index from the corresponding variable is the blue thick plot, while the upper red line and lower black curve in all plots, correspond to the values $n_s=0.9693$ and $n_s=0.9595$, which are the allowed
range of values of  $n_s$, based on the Planck constraints (\ref{planckdata}).}\label{plot1}
\end{figure}

\begin{figure}[h]
\centering
\includegraphics[width=18pc]{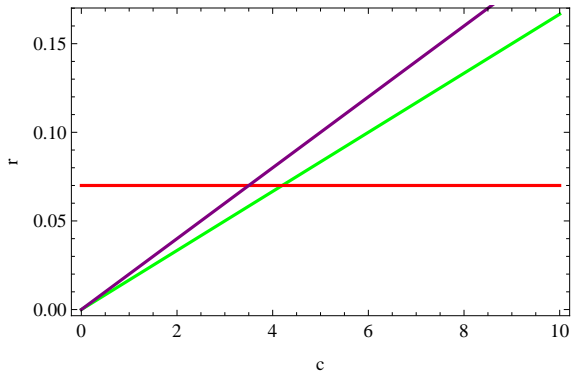}
\caption{The scalar-to-tensor ratio, as a function
of the parameter $c$. The green curve corresponds to $N=60$
and the dashed purple corresponds to $N=50$. The upper red line
corresponds to the BICEP2/Keck-Array upper limit $r=0.07$.
}\label{plot2}
\end{figure}

In order to illustrate the reconstruction technique from the observational indices for the $f(\phi)R$ theory, we shall assume that the Hubble rate is equal to,
\begin{equation}\label{hubbleratemainassumption}
H(N)=\gamma  N^{-\delta }\, ,
\end{equation}
which as we shall demonstrate in this section, it may lead to an inflationary theory, for which $\ddot{a}(t)>0$, for a wide range of the parameters $\gamma$, $\delta$ and the rest parameters of the resulting theory. Also we assume that the scalar-to-tensor ratio has the following form,
\begin{equation}\label{scalartotensorratiofrphi}
r=\frac{c}{N}\, ,
\end{equation}
where $c>0$, which implies that $g(N)=\frac{c}{N}$ in Eq. (\ref{diffeqtionfrphiscalar}). By combining Eqs. (\ref{scalartotensornfrphi}) and (\ref{scalartotensorratiofrphi}), we obtain the following differential equation,
\begin{equation}\label{diffeqnsfrphiscalartotensor}
16 \left(\frac{f'(N)}{2 f(N)}-\frac{H'(N)}{H(N)}\right)=\frac{c}{N}\, ,
\end{equation}
which can be analytically solved by using the functional form of $H(N)$ from Eq. (\ref{hubbleratemainassumption}), and the resulting solution $f(N)$ is,
\begin{equation}\label{fnphirphi}
f(N)=C_1 N^{\frac{1}{8} (c-16 \delta )}\, ,
\end{equation}
where $C_1$ is an integration constant. By using Eqs. (\ref{gravieqnsslowrollapprx3}), (\ref{hubbleratemainassumption}) and (\ref{fnphirphi}), the function $\phi (N)$ can be obtained and it reads,
\begin{equation}\label{phinfrphi}
\phi (N)=\frac{4 \sqrt{2} \sqrt{c} \sqrt{C_1} N^{\frac{1}{16} (c-16 \delta +8)}}{c-16 \delta +8}+C_2\, ,
\end{equation}
where $C_2$ is an integration constant. Then, by combining Eqs. (\ref{slowrollindicesfrphin}), (\ref{hubbleratemainassumption}) (\ref{fnphirphi}) and (\ref{phinfrphi}), we obtain the slow-roll indices as functions of the $e$-foldings number, which are,
\begin{align}\label{slowrollindicesanalyticformfrphi}
& \epsilon_1(N)=\frac{\delta }{N},\,\,\,\epsilon_3(N)=\frac{1}{16} (c-16 \delta ) N^{\frac{1}{8} (16 \delta -c)+\frac{1}{8} (c-16 \delta )-1},\\ \notag &\epsilon_2(N)=\frac{2 \sqrt{2} \delta ^2 N^{-\frac{c}{16}+\delta -\frac{3}{2}}}{\sqrt{c} \sqrt{C_1}}+\frac{c-16 \delta -8}{16 N}\, ,
\end{align}
and the corresponding spectral index of primordial curvature perturbations are,
\begin{equation}\label{spectralindexanalyticformfrphi}
n_s=1-\frac{4 \sqrt{2} \delta ^2 N^{-\frac{c}{16}+\delta -\frac{3}{2}}}{\sqrt{c} \sqrt{C_1}}+\frac{-\frac{c}{8}-2 \delta +1}{N}
\end{equation}
while the scalar-to-tensor ratio is given in Eq. (\ref{scalartotensorratiofrphi}). Now we can investigate the viability of the theory at hand, for various sets of values of the parameters, and we shall confront the theory with the Planck \cite{Ade:2015lrj} constraints of Eq. (\ref{planckdata}) and with the BICEP2/Keck-Array \cite{Array:2015xqh} constraints appearing in Eq. (\ref{scalartotensorbicep2}). As we now demonstrate, the viability of the theory can occur for various sets of values for the free parameters. Consider for example the following set of values, $\delta=0.22$, $c=0.001$, $C_1=1$, for which, by choosing $N=60$ $e$-foldings, the spectral index reads $n_s=0.963488$, which is compatible with the Planck constraints (\ref{planckdata}). Also by choosing, $\delta=-0.5$, $c=0.001$, $C_1=0.034$, again for $N=60$ $e$-foldings, we obtain, $n_s=0.966$. Also, the value $c=0.001$ in both cases, for $N=60$, gives the scalar-to-tensor ratio the value $r=0.000166667$, which is compatible with both the Planck and BICEP2/Keck-Array constraints. It is worth investigating the parameter space in more detail, and a simple analysis reveals that the spectral index is crucially affected by the parameters $\delta$, $c$, and $C_1$, and more importantly, it is independent of $\gamma$. In order to have a concrete idea for the parameter space, in Fig. \ref{plot1} we plot the behavior of the spectral index $n_s$ as a function of the parameter $c$ (upper left plot), for $N=60$, $\delta=0.22$, $\gamma=30$, $C_1=1$, as a function of $\delta$ (upper right plot) for $N=60$, $c=0.001$, $\gamma=30$, $C_1=1$, and as a function of $C_1$ (bottom plot) for $N=60$, $\delta=0.22$, $\gamma=30$, $c=0.001$. In all the plots, the dependence of the spectral index from the corresponding variable is the blue thick plot, while the upper red line and lower black curve in all plots, correspond to the values $n_s=0.9693$ and $n_s=0.9595$, which are the allowed
range of values of  $n_s$, based on the Planck constraints (\ref{planckdata}). As it can be seen, the compatibility with the Planck data can be achieved for a wide range of values of the parameters.
Also in Fig. \ref{plot2} we plot the behavior of the scalar-to-tensor ratio $r$, as a function of the parameter $c$, for $N=60$ (green curve) and for $N=50$ (purple curve). In all the plots, the flat red line indicates the upper limit constraint from the BICEP2/Keck-Array data (\ref{scalartotensorbicep2}).
Hence, the viability of the theory is guaranteed for the present $f(\phi)R$ theory. Now what remains is to find the scalar potential, and also to demonstrate that the cosmology (\ref{hubbleratemainassumption}) indeed describes an inflationary evolution with $\ddot{a}(t)>0$. Now, let us calculate the scalar potential  $V(\phi)$  in the slow-roll approximation, by using Eq. (\ref{gravieqnsslowrollapprx12}). So by inverting $\phi (N)$ in Eq. (\ref{phinfrphi}), we obtain,
\begin{equation}\label{nphifrphi}
N(\phi)=2^{-\frac{40}{c-16 \delta +8}} \left(\frac{c-16 \delta +8}{\sqrt{c} \sqrt{C_1}}\right)^{\frac{16}{c-16 \delta +8}} (\phi -C_2)^{\frac{16}{c-16 \delta +8}}\, .
\end{equation}
So by substituting Eq. (\ref{nphifrphi}) in Eq. (\ref{gravieqnsslowrollapprx12}), we obtain the approximate form of the potential in the slow-roll approximation,
\begin{equation}\label{potentiafrphifinal}
V(\phi)=3 \gamma ^2 C_1 2^{-\frac{5 (c-32 \delta )}{c-16 \delta +8}} \left(\frac{c-16 \delta +8}{\sqrt{c} \sqrt{C_1}}\right)^{\frac{2 (c-32 \delta )}{c-16 \delta +8}} (\phi -C_2)^{\frac{2 (c-32 \delta )}{c-16 \delta +8}}\, .
\end{equation}

In the model of $f(\phi)R$ gravity we investigated, there appear four independent free parameters, namely $\delta$, $c$, $C_1$ and $C_2$, and also the $e$-foldings number $N$. The most important from a phenomenological point of view are $c$ and $\delta$. The parameter $c$ enters in the expression of the scalar-to-tensor ratio in Eq. (\ref{scalartotensorratiofrphi}), and of course the $e$-foldings number $N$, which for phenomenological reasons it is required to have values in the interval $N=[50,60]$, in order to produce enough inflation. From the rest of the parameters, $\delta$ is essential in the theory, since the solution for the Hubble rate as function of the $e$-foldings number is important. We need to note that these free parameters are necessary in the reconstruction method we used, since we require from a theory to produce a specific form of the scalar-to-tensor ratio, so the parameter $c$ is like the parameter $\alpha$ in the $\alpha$-attractors theory \cite{alphaattractors}, so it is an inherent constant of the theory. The parameters $C_1$ and $C_2$ are simply integration constants introduced by solving the dynamical equations. Finally, the parameter $\delta$ controls the way  that the Hubble rate scales as a function of $N$, so it will enter the final expression of the Hubble rate as a function of the cosmic time, see Eq. (\ref{hubbleratecosmictime}) below. Hence, only the parameters $\delta$ and $c$ affect the Hubble rate, but all the parameters contribute to the physical observable quantities $n_s$ and $r$. Nevertheless, $C_1$ and $C_2$ merely depend on the initial conditions of the solutions.

\begin{figure}[h]
\centering
\includegraphics[width=18pc]{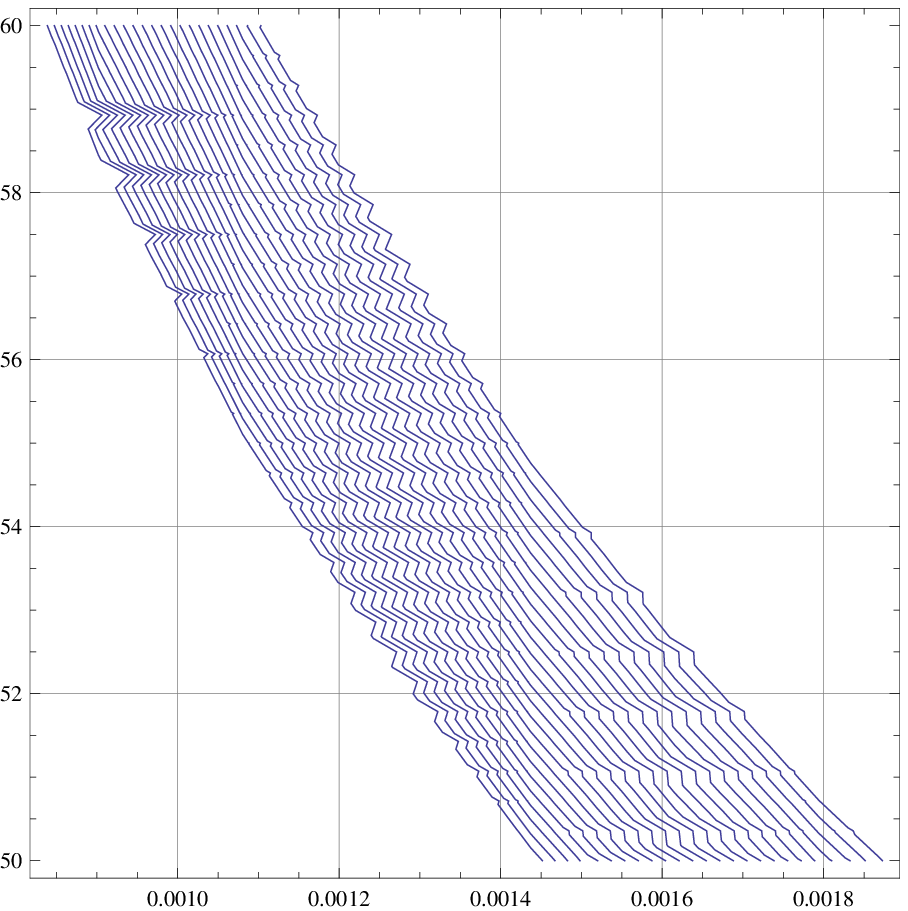}
\includegraphics[width=18pc]{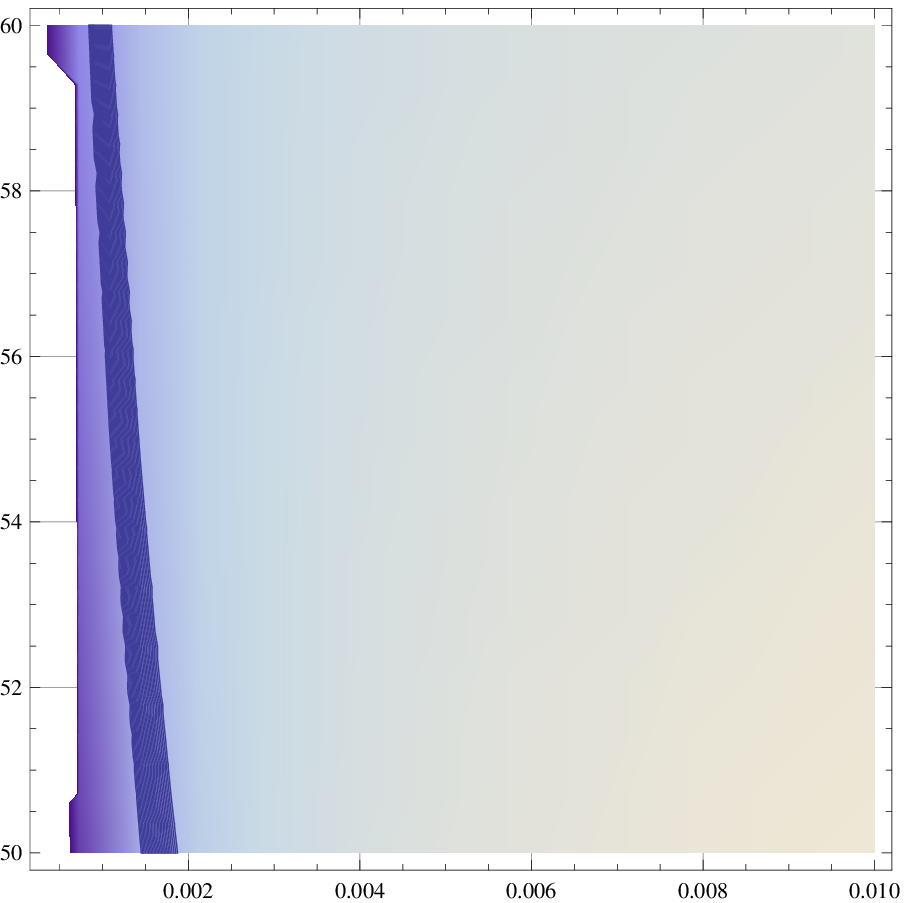}
\caption{Contour plots of the spectral index and of the scalar-to-tensor ratio as functions of the free parameters $c$ and $N$, for $n_s=[0.9595,0.9693]$ and $r<0.07$, and for fixed $\delta$ and $C_1$. Both the conditions are satisfied by the blue curves (left plot) and by the dark blue region (right plot), for a wide range of the parameters $N$ and $c$.}\label{controurdensity}
\end{figure}

\begin{figure}[h]
\centering
\includegraphics[width=18pc]{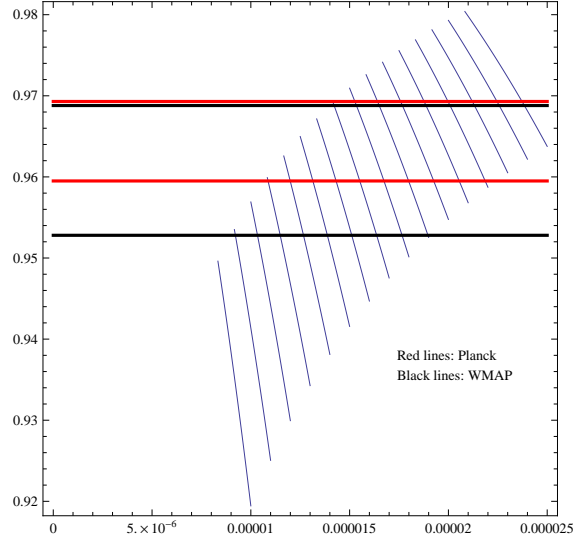}
\caption{The spectral index and the scalar-to-tensor ratio parametric plots for $N=50-60$, for various values of the parameter $c$ and for fixed $\delta$ and $C_1$. As it can be seen, he observational indices may become simultaneously compatible with both the Planck constraints (red lines) and the WMAP constraints (black lines).}\label{parametplots}
\end{figure}

In the analysis we performed so far, we discussed the behavior of the observational indices as functions of the free parameters, and we demonstrated that these can be compatible with the Planck data for a wide range of the free parameters. However we did not discuss the correlation between the spectral index and the scalar-to-tensor ratio. Indeed, this study is important since it is necessary to verify that the observational indices become simultaneously compatible with the observational data, for the same set of the free variables. In order to address this issue, in Fig. \ref{controurdensity}, we plot the contour plots of the spectral index and of the scalar-to-tensor ratio as functions of the $e$-foldings number in the interval $N=[50,60]$ and of the parameter $c$ in the interval $r=[0.002,0.07]$, for fixed $\delta$ and $C_1$ as in the previous examples.
The values of the spectral index that are used for the contour plot lie in the interval $n_s=[0.9595,0.9693]$ which correspond to the Planck constraints \cite{Ade:2015lrj}, and for the scalar-to-tensor ratio for $r<0.07$. As it can be seen in both plots of Fig. \ref{controurdensity}, the blue lines in the left plot, and the dark blue region in the right plot, correspond to the allowed values of the observational indices, when these are constrained from the Planck data. Hence, the spectral index $n_s$ and the scalar-to-tensor ratio can be simultaneously compatible with the observational data, for a wide range of the parameters $N$ and $c$. The same analysis can be carried out for the rest of the free parameters, however we omit it for brevity. Also the resulting behavior for $n_s$ is also compatible with the upper bound of the WMAP constraints \cite{wmap} (the Planck constraints further the spectral index from below), which are,
\begin{equation}\label{wmapconstraints}
n_s=0.9608\pm 0.008\,\,\,(95\%\,\,\mathrm{CL}) .
\end{equation}
With regards to the scalar-to-tensor ratio, the results are compatible with the constraints on it by the WMAP \cite{wmap}, which are,
\begin{equation}\label{wmapconstraints1}
r<0.13\,\,\,(95\%\,\,\mathrm{CL}) .
\end{equation}
It is worth also providing some plots of the spectral index and of the scalar-to-tensor ratio, as functions of $c$ and $N$. In Fig. \ref{parametplots} we plot the observational indices $n_s$ and $r$ as functions of the $e$-foldings $N$, for various values of $c$, and for fixed $\delta$ and $C_1$ as in previous cases. Again, the upper red and lower red lines, indicate the upper and lower Planck constraints on the spectral index, and the upper and lower black indicate the upper and lower constraints of the spectral index coming from WMAP \cite{wmap}.
As it can be seen in Fig. (\ref{parametplots}), $n_s$ and $r$ are simultaneously compatible with the Planck data and also with the WMAP data.

At this point, we shall investigate whether the cosmological evolution (\ref{hubbleratemainassumption}) produces an inflationary cosmology with $\ddot{a}(t)>0$. The analysis is important, since we shall use the same Hubble rate for the mimetic $F(R)$ gravity case in the next section. By using the definition of the $e$-foldings number,
\begin{equation}\label{nfoldingsdef}
N=\int H(t)\mathrm{d}t\, ,
\end{equation}
and by inverting the resulting $N(t)$, the Hubble rate as a function of the cosmic time is,
\begin{equation}\label{hubbleratecosmictime}
H(t)=\gamma  (\delta +1)^{-\frac{\delta }{\delta +1}} \gamma ^{-\frac{\delta }{\delta +1}} t^{-\frac{\delta }{\delta +1}}\, .
\end{equation}
The corresponding scale factor is,
\begin{equation}\label{scalefactorhubblemain}
a(t)=C_6 e^{(\delta +1)^{1-\frac{\delta }{\delta +1}} \gamma ^{\frac{1}{\delta +1}} t^{\frac{1}{\delta +1}}}\, ,
\end{equation}
where $C_6$ is an integration constant, so the second derivative of the scale factor is,
\begin{equation}\label{secondderivetiveddot}
\ddot{a}(t)=C_6 (\delta +1)^{-\frac{2 \delta }{\delta +1}-1} \gamma ^{\frac{1}{\delta +1}} t^{\frac{1}{\delta +1}-2} e^{(\delta +1)^{\frac{1}{\delta +1}} \gamma ^{\frac{1}{\delta +1}} t^{\frac{1}{\delta +1}}} \left((\delta +1) \gamma ^{\frac{1}{\delta +1}} t^{\frac{1}{\delta +1}}-\delta  (\delta +1)^{\frac{\delta }{\delta +1}}\right)\, .
\end{equation}
If the parameter $\delta$ takes negative values, the function $\ddot{a}$ is always positive, however, if $\delta$ takes positive values, $\gamma$ has take very large values, however the parameter $\gamma$ does not affect the observational indices, as we demonstrated earlier. This analysis will hold also true for the mimetic $F(R)$ gravity case, where we shall assume that the Hubble rate is also given by (\ref{hubbleratemainassumption}).

%%%%%%%%%%%%%%%%%%%%%%%%%%%%%%%%%%%%%%%%%%%%%%%%%%%%%%%%%%%%%%%%%%%%%%%%%%%%%%%%%%%%%%%%%%%%%%%%%%%%%%%%%%%%%%%

\begin{figure}[h]
\centering
\includegraphics[width=18pc]{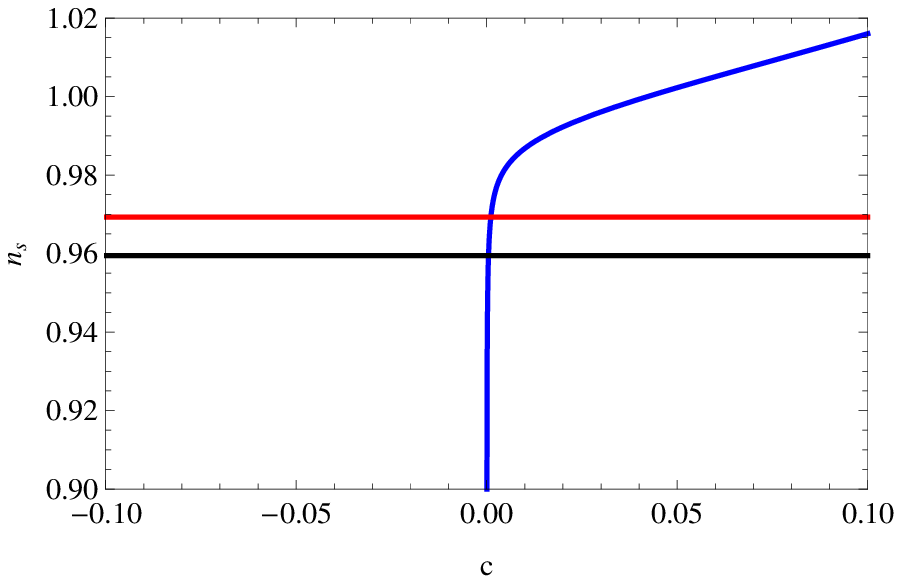}
\includegraphics[width=18pc]{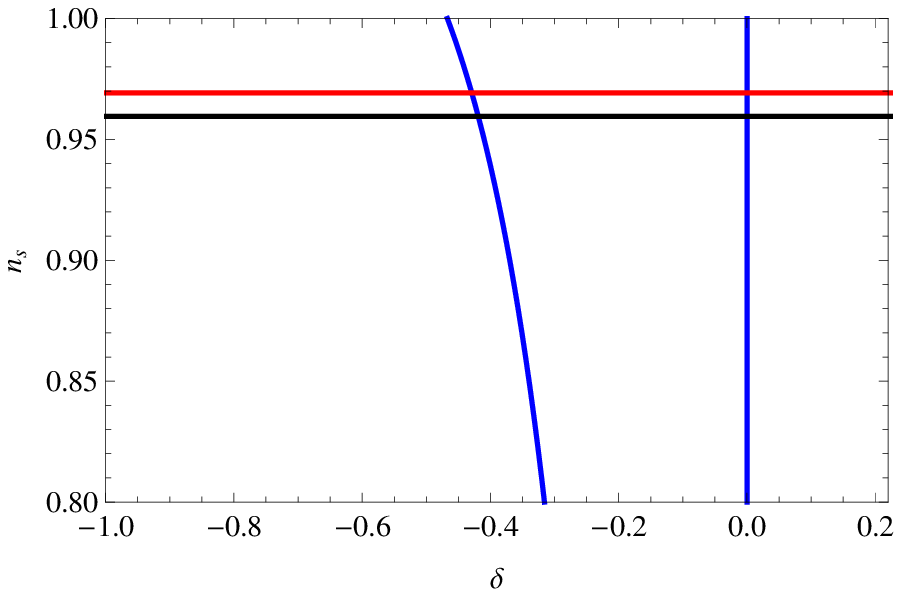}
\includegraphics[width=18pc]{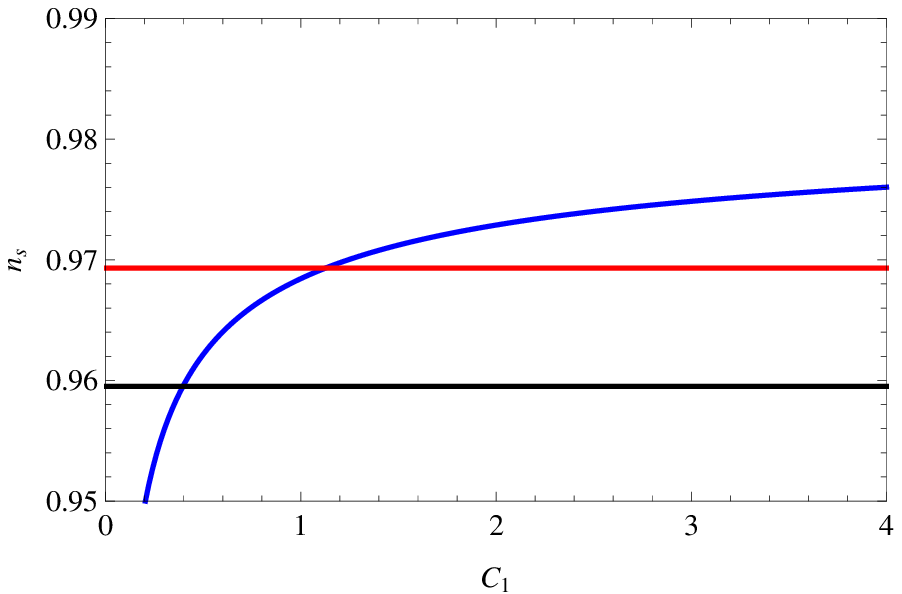}
\includegraphics[width=18pc]{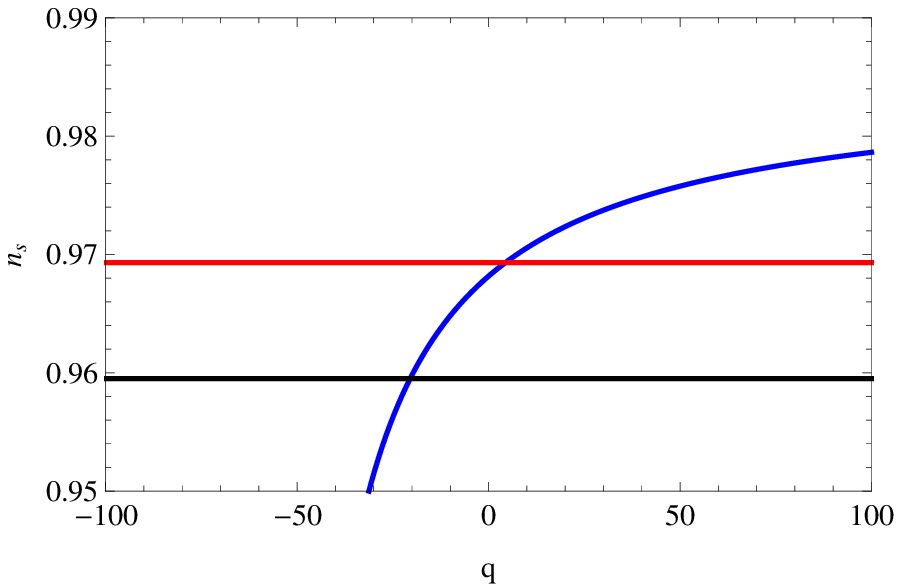}
\caption{The behavior of the spectral index $n_s$ as a function of the $c$ (upper left plot), for $N=60$, $\delta=0.5004$, $C_1=1$, $q=1$, as a function of $\delta$ (upper right plot) for $N=60$, $c=0.001$, $C_1=1$, $q=1$, as a function of $C_1$ (bottom left plot) for $N=60$, $\delta=0.5004$, $c=0.001$, $q=1$ (bottom right plot).}\label{plot3}
\end{figure}

\begin{figure}[h]
\centering
\includegraphics[width=18pc]{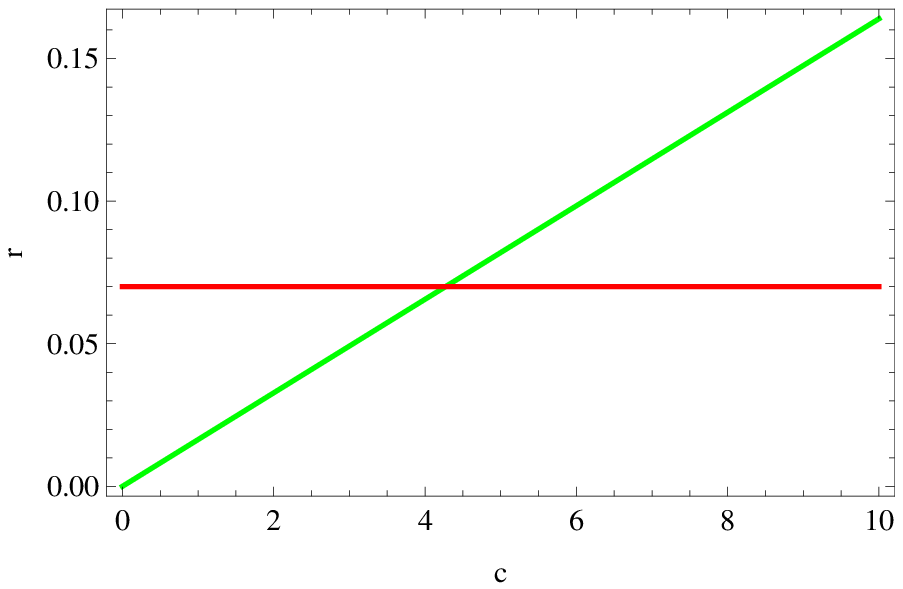}
\includegraphics[width=18pc]{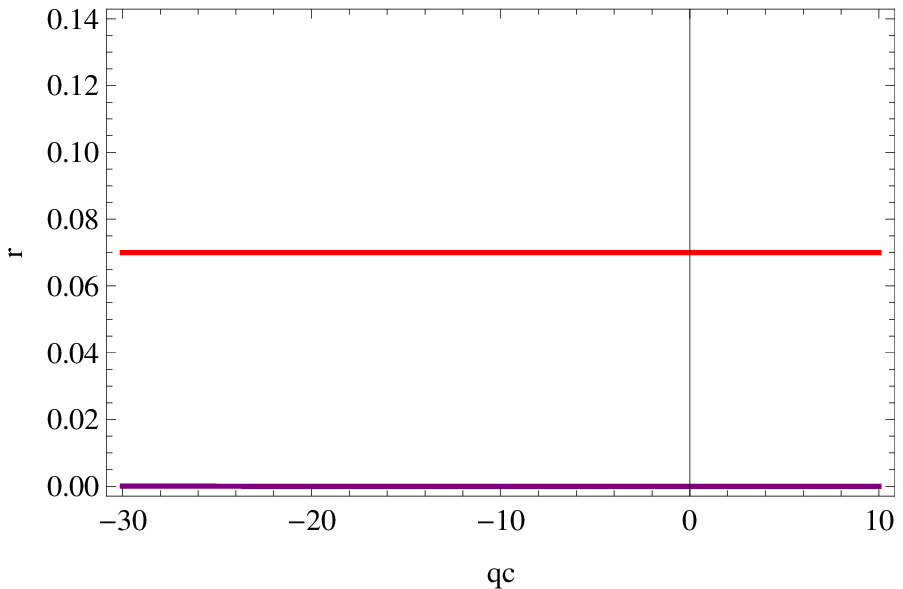}
\caption{The scalar-to-tensor ratio, as a function
of the parameter $c$ (left plot) and as a function of $q$ (right plot).
}\label{plot4}
\end{figure}

Let us now consider another cosmological evolution, in which case the Hubble rate is,
\begin{equation}\label{hubbleratemainassumptiontougianni}
H(N)=\gamma  (N+q)^{-\delta }\, ,
\end{equation}
which as we demonstrate later on section, it leads to the inflationary theory (\ref{hubbleratecosmictime}), for which as we will demonstrate, we have $\ddot{a}(t)>0$. Also we assume that the scalar-to-tensor ratio is equal to,
\begin{equation}\label{scalartotensorratiofrphitougianni}
r=\frac{c}{N+q}\, ,
\end{equation}
where $c>0$ and $q$ is a real number, which implies that $g(N)=\frac{c}{N+q}$ in Eq. (\ref{diffeqtionfrphiscalar}). By combining Eqs. (\ref{scalartotensornfrphi}) and (\ref{scalartotensorratiofrphi}), we obtain,
\begin{equation}\label{diffeqnsfrphiscalartotensortougianni}
16 \left(\frac{f'(N)}{2 f(N)}-\frac{H'(N)}{H(N)}\right)=\frac{c}{N+q}\, ,
\end{equation}
and by solving it, we obtain the function $f(N)$ which is,
\begin{equation}\label{fnphirphitougianni}
f(N)=C_1 (N+q)^{\frac{1}{8} (c-16 \delta )}\, ,
\end{equation}
where $C_1$ is again an integration constant. By using Eqs. (\ref{gravieqnsslowrollapprx3}), (\ref{hubbleratemainassumption}) and (\ref{fnphirphi}), the function $\phi (N)$ reads,
\begin{equation}\label{phinfrphitougianni}
\phi (N)=\frac{4 \sqrt{2} \sqrt{c} \sqrt{C_1} (N+q)^{\frac{c}{16}-\frac{1}{2} (2 \delta +1)+1}}{c-16 \delta +8}+C_2\, ,
\end{equation}
where $C_2$ is an integration constant. Then, by using Eqs. (\ref{slowrollindicesfrphin}), (\ref{hubbleratemainassumption}) (\ref{fnphirphi}) and (\ref{phinfrphi}), the slow-roll indices read,
\begin{align}\label{slowrollindicesanalyticformfrphitougianni}
& \epsilon_1(N)=\frac{\delta }{N+q},\,\,\,\epsilon_3(N)=\frac{1}{16} (c-16 \delta ) (N+q)^{\frac{1}{8} (16 \delta -c)+\frac{1}{8} (c-16 \delta )-1},\\ \notag &\epsilon_2(N)=\frac{(N+q)^{-\frac{c}{16}-2}}{16 \sqrt{2} \sqrt{c} \sqrt{C_1} (c-32 \delta )}\times \\ \notag &
\left(\sqrt{2} c^{5/2} \sqrt{C_1} (N+q)^{\frac{c}{16}+1}+64 c \delta ^2 (N+q)^{\delta +\frac{1}{2}} \sqrt{(N+q)^{2 \delta +1}}-512 \delta ^2 (2 \delta -1) (N+q)^{\delta +\frac{1}{2}} \sqrt{(N+q)^{2 \delta +1}}\right.\\ \notag &
+\left.512 \sqrt{2} \sqrt{c} \sqrt{C_1} \delta  (2 \delta +1) (N+q)^{\frac{c}{16}+1}-16 \sqrt{2} c^{3/2} \sqrt{C_1} (4 \delta +1) (N+q)^{\frac{c}{16}+1}\right)
\, ,
\end{align}
and the spectral index can be easily calculated and it reads,,
\begin{align}\label{spectralindexanalyticformfrphitougianni}
& n_s=\frac{1}{8 \sqrt{c} \sqrt{C_1} (c-32 \delta )^2}(N+q)^{-\frac{c}{16}-2}\left(c^{7/2} \sqrt{C_1} (N+q)^{\frac{c}{16}+1} \left(-3 N (N+q)^{2 \delta }-3 q (N+q)^{2 \delta }+2\right)\right.\\ \notag &
-32 \sqrt{2} c^2 \delta ^2 (N+q)^{\delta +\frac{1}{2} (2 \delta +1)+\frac{1}{2}}-8192 \sqrt{2} \delta ^3 (2 \delta -1) (N+q)^{\delta +\frac{1}{2} (2 \delta +1)+\frac{1}{2}}\\ \notag &
\left.-22 \delta -6 q (N+q)^{2 \delta }+18 \delta  N (N+q)^{2 \delta }+18 \delta  q (N+q)^{2 \delta }\right)\\ \notag &
\left.+N \left(12 \delta ^2 (N+q)^{2 \delta }+3 (N+q)^{2 \delta }+\delta  \left(8-12 (N+q)^{2 \delta }\right)\right)\right)\\ \notag &\left.\left.+N \left(36 \delta ^2 (N+q)^{2 \delta }+3 (N+q)^{2 \delta }+\delta  \left(8-24 (N+q)^{2 \delta }\right)\right)\right)\right)\, ,
\, ,
\end{align}
and the scalar-to-tensor ratio is given in Eq. (\ref{scalartotensorratiofrphitougianni}). Having the observational indices at hand, we can investigate the viability of the present theory, by using various values for the parameters. As we show, the viability easily comes, consider for example the set of values $\delta=0.5004$, $c=0.001$, $C_1=1$, $q=1$, the spectral index reads, $n_s=0.96844$, and the scalar-to-tensor ratio is $r=0.000015873$, which are compatible with the Planck constraints (\ref{planckdata}) and the BICEP2/Keck-Array data (\ref{scalartotensorbicep2}). In Fig. \ref{plot3}, we plot the behavior of the spectral index $n_s$ as a function of the $c$ (upper left plot), for $N=60$, $\delta=0.5004$, $C_1=1$, $q=1$, as a function of $\delta$ (upper right plot) for $N=60$, $c=0.001$, $C_1=1$, $q=1$, as a function of $C_1$ (bottom left plot) for $N=60$, $\delta=0.5004$, $c=0.001$, $q=1$ (bottom right plot). In all the plots, the spectral index dependence is the blue thick plot, while the upper red line and lower black curve correspond to the values $n_s=0.9693$ and $n_s=0.9595$. In this case too, the compatibility with the Planck data is achieved for a wide range of values of the parameters.
Moreover, in Fig. \ref{plot4} we plot the scalar-to-tensor ratio $r$, as a function of the parameter $c$, for $N=60$, $q=1$ (left plot) and for $N=60$, $c=0.001$ (right plot). As in the previous scenario, the red line indicates BICEP2/Keck-Array constraint.
Hence, in this case the viability of the theory is guaranteed for a wide range of the parameters values. Also it is easy to show that the resulting potential is identical with the one appearing in Eq. (\ref{potentiafrphifinal}) In addition, by using (\ref{nfoldingsdef}), we obtain the function $N(t)$, which is,
\begin{equation}\label{ntfrphi}
N(t)=((\delta +1) (C_5+\gamma  t))^{\frac{1}{\delta +1}}-q\, ,
\end{equation}
where $C_5$ is an integration constant, and by choosing $C_5=0$, the Hubble rate is identical to the one appearing in Eq. (\ref{hubbleratecosmictime}). Hence by choosing the value of $\gamma\gg 1$, we can obtain $\ddot{a}>0$\footnote{Recall that the observational indices do not depend on the parameter $\gamma$.}

\section{Reconstruction of Viable Inflationary Mimetic $F(R)$ Gravity}

In this section we shall consider the realization of the viable cosmological evolutions of the previous section, in the context of another popular modified gravity, the mimetic $F(R)$ gravity. In the context of mimetic $F(R)$ gravity, the conformal symmetry is not violated, but actually it is a conserved internal degree of freedom \cite{Chamseddine:2013kea}. More importantly, it has been recently shown that the mimetic gravity contains ghosts \cite{Takahashi:2017pje,Nojiri:2017ygt}, and it is possible to formulate it in such a way, so that the resulting theory is ghost free \cite{Nojiri:2017ygt}. Let us recall in brief, the theoretical framework of mimetic $F(R)$ gravity, which we shall extensively use in the rest of this section. In mimetic gravity, there are two metrics, the physical metric $g_{\mu \nu}$ and the auxiliary metric $\hat{g}_{\mu \nu}$, and by using an auxiliary scalar field $\phi$, it is possible to express the physical metric tensor in terms of the auxiliary metric $\hat{g}_{\mu \nu}$ and the scalar field $\phi$, in the following way,
\begin{equation}\label{metrpar}
g_{\mu \nu}=-\hat{g}^{\mu \nu}\partial_{\rho}\phi \partial_{\sigma}\phi
\hat{g}_{\mu \nu}\, .
\end{equation}
Then, from Eq. (\ref{metrpar}), it follows that,
\begin{equation}\label{impl1}
g^{\mu \nu}(\hat{g}_{\mu \nu},\phi)\partial_{\mu}\phi\partial_{\nu}\phi=-1\,
.
\end{equation}
Under the Weyl transformation $\hat{g}_{\mu \nu}=e^{\sigma (x)}g_{\mu \nu}$, the parametrization of the metric (\ref{metrpar}), is invariant, and in effect, the auxiliary metric $\hat{g}_{\mu
\nu}$ does not appear in the resulting gravitational action of the theory. In the following we shall assume a flat FRW metric of the form (\ref{JGRG14}), in which case the scalar curvature has the form $R=6\left (\dot{H}+2H^2 \right )$, with $H$ being the Hubble rate, and also we shall make the assumption that the scalar field depends only on the cosmic time. The mimetic $f(R)$ gravity vacuum gravitational action has the following form \cite{Nojiri:2014zqa},
\begin{equation}\label{actionmimeticfraction}
S=\int \mathrm{d}x^4\sqrt{-g}\left ( f\left(R(g_{\mu \nu})\right
)-V(\phi)+\lambda \left(g^{\mu \nu}\partial_{\mu}\phi\partial_{\nu}\phi
+1\right)\right )\, ,
\end{equation}
The functions $V(\phi)$ and $\lambda (t)$, are the scalar potential of the auxiliary scalar field, and the Lagrange multiplier of the theory. By using the Lagrange multiplier, the mimetic constraint can be realized, as we now demonstrate. By varying the gravitational action (\ref{actionmimeticfraction}) with respect to the physical metric $g_{\mu \nu}$, we obtain the following gravitational equations of motion,
\begin{align}\label{aeden}
& \frac{1}{2}g_{\mu \nu}f(R)-R_{\mu
\nu}F(R)+\nabla_{\mu}\nabla_{\nu}F(R)-g_{\mu \nu}\square F(R)\\ \notag &
\frac{1}{2}g_{\mu \nu}\left (-V(\phi)+\lambda \left( g^{\rho
\sigma}\partial_{\rho}\phi\partial_{\sigma}\phi+1\right) \right )-\lambda
\partial_{\mu}\phi \partial_{\nu}\phi =0 \, ,
\end{align}
where $F=\frac{\partial f}{\partial R}$. Additionally, by varying the gravitational action of Eq. (\ref{actionmimeticfraction}), with respect to the auxiliary scalar field $\phi$, we obtain the following equation,
\begin{equation}\label{scalvar}
-2\nabla^{\mu} (\lambda \partial_{\mu}\phi)-V'(\phi)=0\, ,
\end{equation}
where the ``prime'' in this instance indicates differentiation with respect to the auxiliary scalar $\phi$. Also, by varying the gravitational action with respect to $\lambda$, we obtain the following equation,
\begin{equation}\label{lambdavar}
g^{\rho \sigma}\partial_{\rho}\phi\partial_{\sigma}\phi=-1\, .
\end{equation}
By assuming that the physical metric is described by the FRW metric (\ref{JGRG14}), and also that the auxiliary scalar depends only on the cosmic time, the gravitational equations of motion (\ref{aeden}), (\ref{scalvar})
and (\ref{lambdavar}) take the following form,
\begin{equation}\label{enm1}
-f(R)+6(\dot{H}+H^2)F(R)-6H\frac{\mathrm{d}F(R)}{\mathrm{d}t}-\lambda
(\dot{\phi}^2+1)+V(\phi)=0\, ,
\end{equation}
\begin{equation}\label{enm2}
f(R)-2(\dot{H}+3H^2)+2\frac{\mathrm{d}^2F(R)}{\mathrm{d}t^2}+4H\frac{\mathrm{d}F(R)}{\mathrm{d}t}-\lambda (\dot{\phi}^2-1)-V(\phi)=0\, ,
\end{equation}
\begin{equation}\label{enm3}
2\frac{\mathrm{d}(\lambda \dot{\phi})}{\mathrm{d}t}+6H\lambda
\dot{\phi}-V'(\phi)=0\, ,
\end{equation}
\begin{equation}\label{enm4}
\dot{\phi}^2-1=0\, ,
\end{equation}
where the ``dot'' denotes differentiation with respect to the cosmic time $t$. Moreover, only for this instance, the prime in Eq. (\ref{enm3}) denotes differentiation with respect to $\phi$. From the mimetic constraint (\ref{enm3}), it easily follows that the identification of the auxiliary scalar field with the cosmic time is possible, that is $\phi=t$. The latter identification is also possible in the context of Einstein-Hilbert mimetic gravity, see Ref. \cite{Chamseddine:2013kea}. By using the identification $\phi=t$, the gravitational equation (\ref{enm2}) can be cast in the following form,
\begin{equation}\label{sone}
f(R)-2(\dot{H}+3H^2)+2\frac{\mathrm{d}^2F(R)}{\mathrm{d}t^2}+4H\frac{\mathrm{d}F(R)}{\mathrm{d}t}-V(t)=0\, ,
\end{equation}
and in effect, the mimetic scalar potential $V(\phi=t)$ in terms of the Hubble rate and the $f(R)$ gravity of the theory, is equal to,
\begin{equation}\label{scalarpot}
V(\phi=t)=2\frac{\mathrm{d}^2F(R)}{\mathrm{d}t^2}+4H\frac{\mathrm{d}F(R)}{
\mathrm{d}t}+f(R)-2(\dot{H}+3H^2)\, .
\end{equation}
In effect, if we combine Eqs. (\ref{scalarpot}) and (\ref{enm1}), we may obtain the analytic form of the Lagrange multiplier function $\lambda (t)$ in terms of the Hubble rate and the $F(R)$ gravity,
\begin{equation}\label{lagrange}
\lambda (t)=-3 H \frac{\mathrm{d}F(R)}{\mathrm{d}t}+3
(\dot{H}+H^2)-\frac{1}{2}f(R)\, .
\end{equation}
By using the above equations and expressing these in terms of the $e$-foldings number, we will have a reconstruction method at hand which will enable us to generate viable inflationary cosmological evolutions in the context of mimetic $F(R)$ gravity. We shall explain in more detail the reconstruction method from the observational indices, but let us first recall the essential features of inflationary dynamics in the context of mimetic $F(R)$ gravity. For more details on these issues consult the review \cite{reviews1} and also Ref. \cite{Nojiri:2016vhu}.

The inflationary dynamics of mimetic $F(R)$ gravity in the Jordan frame is captured by the slow-roll indices $\epsilon_i$, $i=1,...,4$. We shall assume that the slow-roll conditions holds true, in which case the following conditions hold true,
\begin{equation}
\label{slowrollerarealtions}
\dot{H}\ll H^2\, ,\quad \ddot{H}\ll H\dot{H}\, ,
\end{equation}
with $H$ being the Hubble rate, and in effect the resulting slow-roll indices are simplified to a great extent, as we show shortly. With the mimetic constraint (\ref{enm4}) holding true, which results to $\phi=t$ as we showed, the mimetic $F(R)$ gravity theory, with Lagrange multiplier and scalar potential, can be regarded as a generalized $F(R,\phi)$ scalar-tensor theory, with the action being cast as follows,
\be
\label{MF1sc}
S = \int \sqrt{-g} \left\{ \frac{f(R)}{2\kappa^2} + \lambda (\phi) \partial_\mu \phi \partial^\mu \phi
+ \lambda(\phi ) - V(\phi) \right\} \, ,
\ee
in which case the kinetic term is $\omega (t)=-2 \lambda (t )$, and the corresponding generalized scalar potential $\mathcal{V}(t)$ has the form $\mathcal{V}(t)=\lambda (t)-V(t)$. In the case at hand, the slow-roll indices take the following form \cite{reviews1},
\begin{equation}
\label{slowrollindices}
\epsilon_1=-\frac{\dot{H}}{H^2}\, ,\quad \epsilon_2=\frac{\ddot{\phi}}{H\dot{\phi}}\, , \quad \epsilon_3=\frac{\dot{F}(R,\phi)}{2HF(R,\phi)}\, , \quad \epsilon_4=\frac{\dot{E}}{2HE}\, ,
\end{equation}
where the function $E(R,\phi)$ has the following form,
\begin{equation}
\label{epsilonfunction}
E(R,\phi)=F(R,\phi)\omega (\phi)+\frac{3\dot{F}(R,\phi)^2}{2\dot{\phi}^2}\, ,
\end{equation}
Due to the mimetic constraint identification $\phi=t$, and also due to the fact that $\omega (t)=-2\lambda (t)$, the slow-roll indices can be cast in the following form,
\begin{equation}
\label{slowrollindicesmim}
\epsilon_1=-\frac{\dot{H}}{H^2}\, ,\quad \epsilon_2=0\, ,\quad \epsilon_3=\frac{\dot{F}(R,\phi)}{2HF(R,\phi)}\, ,\quad \epsilon_4=\frac{\dot{E}}{2HE}\, ,
\end{equation}
and the function $E$ may be expressed as follows,
\begin{equation}
\label{epsilonfunction1}
E(R,\phi)=-2F(R,\phi)\lambda (\phi)+\frac{3\dot{F}(R,\phi)^2}{2}\, .
\end{equation}
Accordingly, for the theory at hand, after calculating the slow-roll indices, the observational indices can be calculated, and we shall be mainly interested in the spectral index of the primordial curvature perturbations and also in the scalar-to-tensor ratio, which are in this case \cite{reviews1},
\begin{equation}
\label{obsindices}
n_s\simeq 1-4\epsilon_1-2\epsilon_2+2\epsilon_3-2\epsilon_4\, ,\quad r\simeq 16(\epsilon_1+\epsilon_3)\, .
\end{equation}
The reconstruction method can be revealed once the above quantities can be expressed in terms of the $e$-foldings number. Apparently, the expression for the scalar-to-tensor ratio corresponding to the mimetic $f(R)$ gravity is different in comparison to the standard $f(R)$ gravity. In the Appendix C we prove the relation $r\simeq 16(\epsilon_1+\epsilon_3)$, which holds true only if the slow-roll approximation is assumed.

So let us express the slow-roll indices in terms of the $e$-foldings number, so the slow-roll indices (\ref{slowrollindicesmim}) for the mimetic $F(R)$ gravity are,
\begin{align}\label{slowrollexplicitmimeticfrnfoldings}
& \epsilon_1=-\frac{H'(N)}{H(N)},\,\,\,\epsilon_2=0,\,\,\, \epsilon_3=\frac{F'(N)}{2 F(N)},\\ \notag &
\epsilon_4=\frac{-2 \lambda (N) F'(N)+3 F'(N) \left(H(N)^2 F''(N)+H(N) H'(N)^2\right)-2 F(N) \lambda '(N)}{2 \left(\frac{3}{2} H(N)^2 H'(N)^2-2 F(N) \lambda (N)\right)}\, ,
\end{align}
where the prime indicates differentiation with respect to the $e$-foldings number. The corresponding spectral index of the primordial curvature perturbation is equal to,
\begin{equation}\label{spectralindexgeneralmimeticform}
n_s=1+\frac{F'(N)}{F(N)}-\frac{-2 \lambda (N) F'(N)+3 F'(N) \left(H(N)^2 F''(N)+H(N) H'(N)^2\right)-2 F(N) \lambda '(N)}{\frac{3}{2} H(N)^2 H'(N)^2-2 F(N) \lambda (N)}+\frac{4 H'(N)}{H(N)}\, ,
\end{equation}
while the scalar-to-tensor ratio is,
\begin{equation}\label{scalartotensoratiogeneralmimetic}
r=16 \left(\frac{F'(N)}{2 F(N)}-\frac{H'(N)}{H(N)}\right)\, .
\end{equation}
Now the reconstruction technique is based on the slow-roll assumption, in which case the scalar curvature is approximately equal to,
\begin{equation}\label{appproxscalarcurv}
R\simeq 12H^2(N)\, ,
\end{equation}
so by specifying the Hubble rate and also by assuming that the scalar-to-tensor ratio has a particular desirable  form $r=g(N)$, from Eq. (\ref{scalartotensoratiogeneralmimetic}) we obtain the following differential equation,
\begin{equation}\label{differentialequationgeneralmimetic}
16 \left(\frac{F'(N)}{2 F(N)}-\frac{H'(N)}{H(N)}\right)=g(N)\, ,
\end{equation}
which when solved, the function $F(N)$ is obtained. Then by inverting Eq. (\ref{appproxscalarcurv}), we obtain the function $N=N(R)$, so by substituting in the resulting $F(N)$ gravity, we obtain the function $F(N(R))$. Accordingly, by integrating once with respect to the Ricci scalar, we obtain the function $f(R)$. After this step, it is possible to calculate the functions $\lambda (N)$ and $V(N)$ from Eqs. (\ref{scalarpot}) and (\ref{lagrange}). These equations in terms of the $e$-foldings number $N$ become,
\begin{align}\label{breakscalarpotentialefoldings}
& V(N)=\frac{f(N)}{2}-F(N) \left(H(N) H'(N)+3 H(N)^2\right)+6F'(N)\left(8 H(N)^3 H'(N)+4 H(N)^2 H'(N)^2\right. \\ \notag &
\left.+6 H(N) \left(H''(N)+H(N) H'(N)^2\right)+H(N) \left(H'''(N)+H'(N)^3+2 H(N) H'(N) H''(N)\right)\right)\, ,
\end{align}
\begin{align}\label{breaklagrangemultipl}
& \lambda (N)=\frac{1}{2} \left(-\frac{f(N)}{2}-18 F'(N) \left(4 H(N)^3 H'(N)+H(N) \left(H''(N)+H(N) H'(N)^2\right)\right)+F(N) \left(H(N) H'(N)+3 H(N)^2\right)\right)\\ \notag &
\left.\left.+6 H(N) \left(H''(N)+H(N) H'(N)^2\right)+H(N) \left(H'''(N)+H'(N)^3+2 H(N) H'(N) H''(N)\right)\right)\right)\, .
\end{align}
Having these at hand, the spectral index can be calculated, and the viability of the resulting inflationary theory can be investigated thereafter.

\begin{figure}[h]
\centering
\includegraphics[width=18pc]{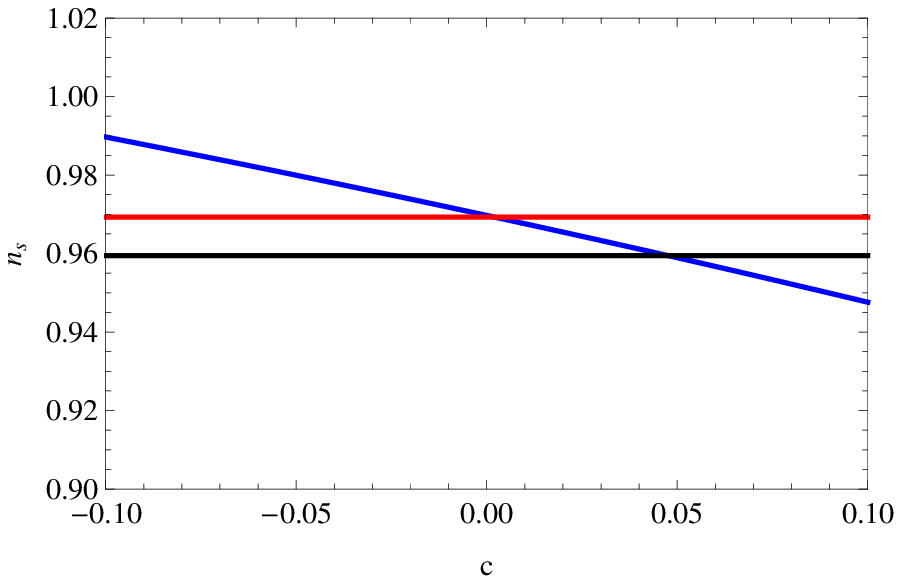}
\includegraphics[width=18pc]{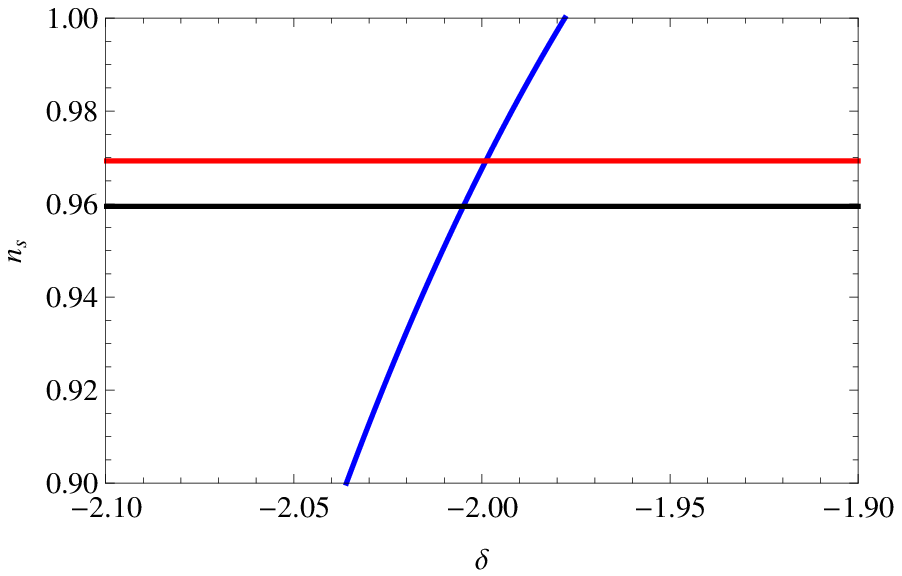}
\includegraphics[width=18pc]{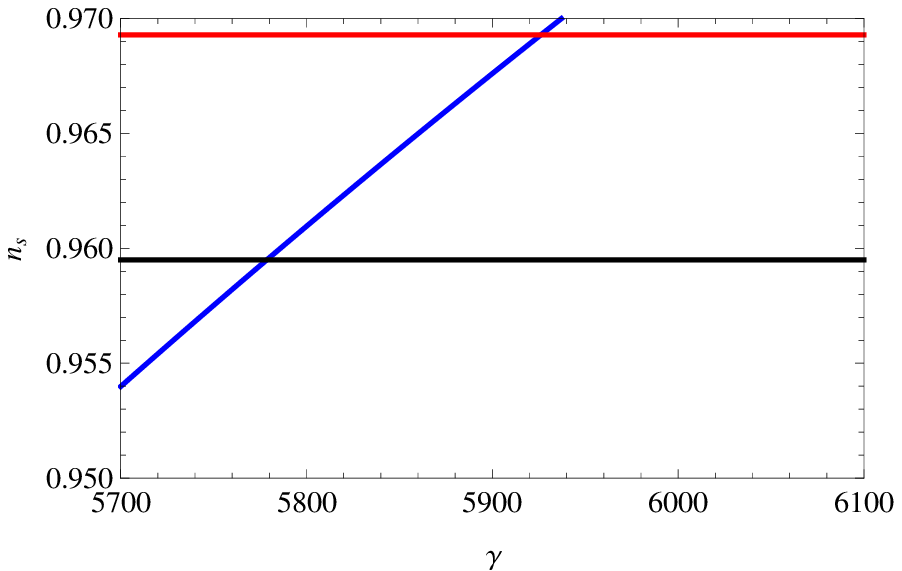}
\caption{The spectral index $n_s$ as a function of the $c$ (upper left plot), for $N=60$, $\delta=-2$, $\gamma=5.9\times 10^3$, $C_1=1$ and $C_2=1$, as a function of $\delta$ (upper right plot), for $c=0.01$, $\gamma=5.9\times 10^3$, $C_1=1$ and $C_2=1$ and finally, as a function of $\gamma$, for $c=0.01$, $\delta=-2$, $C_1=1$ and $C_2=1$ (lower plot).}\label{plot5}
\end{figure}

\begin{figure}[h]
\centering
\includegraphics[width=18pc]{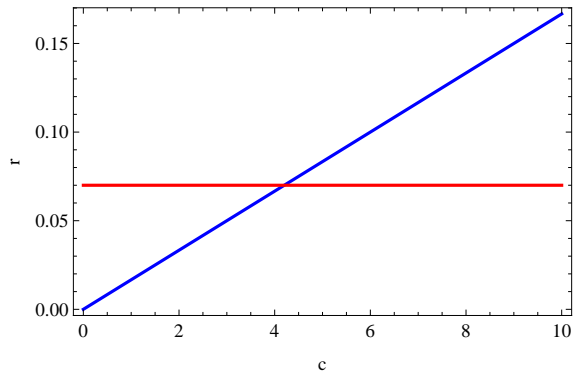}
\caption{The scalar-to-tensor ratio, as a function
of the parameter $c$.
}\label{plot6}
\end{figure}

We shall consider two types of cosmological evolutions, which we used in the previous sections, namely the inflationary cosmologies with the Hubble rates appearing in Eqs. (\ref{hubbleratemainassumption}) and (\ref{hubbleratemainassumptiontougianni}). We start of with the cosmological evolution (\ref{hubbleratemainassumption}), so let us assume that the scalar-to-tensor ratio is,
\begin{equation}\label{scalartotensorratiofrphitougianniextra}
r=\frac{c}{N}\, ,
\end{equation}
with $c>0$, so we have $g(N)=\frac{c}{N}$ in Eq. (\ref{differentialequationgeneralmimetic}), so the resulting differential equation (\ref{differentialequationgeneralmimetic}) becomes,
\begin{equation}\label{differentialequationgeneralmimetic}
16 \left(\frac{F'(N)}{2 F(N)}-\frac{H'(N)}{H(N)}\right)=\frac{c}{N}\, ,
\end{equation}
which can be solved analytically, and the resulting $F(N)$ function is,
\begin{equation}\label{fnphirphitougianniextra}
F(N)=C_1 N^{\frac{1}{8} (c-16 \delta )}\, ,
\end{equation}
with $C_1$ being an integration constant.
The scalar curvature $R$ in terms of $N$ is $R=12 \gamma ^2 N^{-2 \delta }$, so by inverting this, we obtain the function $N(R)=2^{1/\delta } 3^{\frac{1}{2 \delta }} \left(\frac{R}{\gamma ^2}\right)^{-\frac{1}{2 \delta }}$. Hence the function $F(R)$ is equal to,
\begin{equation}\label{frfinalderivative}
F(R)=C_1 2^{\frac{c}{8 \delta }-2} 3^{\frac{c}{16 \delta }-1} \gamma ^{4 \delta -\frac{c}{4}} R^{1-\frac{c}{16 \delta }}\, ,
\end{equation}
and by integrating, we obtain the $f(R)$ gravity which is,
\begin{equation}\label{caseifrgravity}
f(R)=C_2-\frac{R^{2-\frac{c}{16 \delta }} \left(C_1 \delta  2^{\frac{c}{8 \delta }+2} 3^{\frac{c}{16 \delta }-1} \gamma ^{4 \delta -\frac{c}{4}}\right)}{c-32 \delta }\, .
\end{equation}
Having this at hand, the function $f(N)$ easily follows, and it reads,
\begin{equation}\label{fN}
f(N)=C_2-\frac{192 C_1 \delta  \gamma ^{-\frac{2 c \delta +c}{8 \delta }+4 \delta +4} N^{\frac{c}{8}-4 \delta }}{c-32 \delta }\, .
\end{equation}
Accordingly, the mimetic potential can be found from Eq. (\ref{breakscalarpotentialefoldings}), and it reads,
\begin{align}\label{mimeticpotentialofnexplicitform1}
& V(N)=\frac{1}{2} \left(C_2-\frac{192 C_1 \delta  \gamma ^{-\frac{2 c \delta +c}{8 \delta }+4 \delta +4} N^{\frac{c}{8}-4 \delta }}{c-32 \delta }\right)-C_1 \gamma ^{-\frac{(2 \delta +1) (c-16 \delta )}{8 \delta }} N^{\frac{c}{8}-2 \delta } \left(3 \gamma ^2 N^{-2 \delta }-\gamma ^2 \delta  N^{-2 \delta -1}\right)\\ \notag &
\frac{1}{2} \left(C_2-\frac{192 C_1 \delta  \gamma ^{-\frac{2 c \delta +c}{8 \delta }+4 \delta +4} N^{\frac{c}{8}-4 \delta }}{c-32 \delta }\right)-C_1 \gamma ^{-\frac{(2 \delta +1) (c-16 \delta )}{8 \delta }} N^{\frac{c}{8}-2 \delta } \left(3 \gamma ^2 N^{-2 \delta }-\gamma ^2 \delta  N^{-2 \delta -1}\right)\\ \notag &
\left.\gamma  +N^{-\delta } \left(-\gamma ^3 \delta ^3 N^{-3 \delta -3}+2 \gamma ^3 (-\delta -1) \delta ^2 N^{-3 \delta -3}-\gamma  (-\delta -2) (-\delta -1) \delta  N^{-\delta -3}\right)\right)\, ,
\end{align}
and also the Lagrange multiplier $\lambda (N)$ reads,
\begin{equation}\label{lagrangemultiplirer}
\lambda (N)=\frac{1}{64} C_1 (c-16 \delta ) \gamma ^{-\frac{(2 \delta +1) (c-16 \delta )}{8 \delta }} N^{\frac{c}{8}-4 \delta -3} \left(\gamma ^2 \delta  (3 \delta +2)+\left(\delta ^2+3 \delta +2\right) N^{2 \delta }-3 (\delta +1) N^{2 \delta +1}-4 \gamma ^2 N^2-7 \gamma ^2 \delta  N\right)\, .
\end{equation}
Having the analytic form of the scalar potential and of the Lagrange multiplier, we can calculate the slow-roll indices, the analytic form of $\epsilon_1$, $\epsilon_2$ and $\epsilon_3$ is,
\begin{equation}\label{slowrollanalyticforms}
\epsilon_1=\frac{\delta }{N},\,\,\,\epsilon_2=0,\,\,\,\epsilon_3=\frac{1}{16} (c-16 \delta ) N^{\frac{1}{8} (16 \delta -c)+\frac{1}{8} (c-16 \delta )-1}\, ,
\end{equation}
while the slow-roll index $\epsilon_4$ is,
\begin{align}\label{epsilon4analyticmimetic1}
& \epsilon_4=\left(3 C_1 N (c-16 \delta ) N^{\frac{c}{8}-1} \gamma ^{\frac{2 c \delta +c}{8 \delta }} \left(C_1 (c-16 \delta -8) (c-16 \delta ) N^{c/8}+64 \gamma  \delta ^2 N^{\delta }\right)\right.\\ \notag &
\left(3 C_1 N (c-16 \delta ) N^{\frac{c}{8}-1} \gamma ^{\frac{2 c \delta +c}{8 \delta }} \left(C_1 (c-16 \delta -8) (c-16 \delta ) N^{c/8}+64 \gamma  \delta ^2 N^{\delta }\right)\right.\\ \notag &
+4 ((\delta +2) (4 \delta +3)-6 (2 \delta +1) N)))))\times \\ \notag &
\left.\left.(\delta +1) (\delta -3 N+2) N^{2 \delta }+\gamma ^2 \left(\delta  (3 \delta +2)-4 N^2-7 \delta  N\right)\right)\right)^{-1}\, .
\end{align}
Accordingly, by substituting these in Eq. (\ref{obsindices}), the spectral index reads,
\begin{align}\label{spectralindexanalytcmimetic1}
& n_s=-\left(3 c^3 C_1^2 N^{\frac{c}{4}+1} \gamma ^{\frac{2 c \delta +c}{8 \delta }}+96 c \delta  N \gamma ^{\frac{2 c \delta +c}{8 \delta }} \left(8 C_1^2 (3 \delta +1) N^{c/4}+2 \gamma  C_1 \delta  N^{\frac{c}{8}+\delta }-\gamma ^2 \delta  N^{2 \delta }\right)\right.\\ \notag &
-\left(3 c^3 C_1^2 N^{\frac{c}{4}+1} \gamma ^{\frac{2 c \delta +c}{8 \delta }}+96 c \delta  N \gamma ^{\frac{2 c \delta +c}{8 \delta }} \left(8 C_1^2 (3 \delta +1) N^{c/4}+2 \gamma  C_1 \delta  N^{\frac{c}{8}+\delta }-\gamma ^2 \delta  N^{2 \delta }\right)\right.\\ \notag &
\left.+\gamma ^2 \left(-\delta  (2 \delta +3) (3 \delta +2)+4 N^3+(15 \delta +4) N^2+\delta  (11 \delta +12) N\right)\right)\\ \notag & +256\delta \left(-3 \delta  N \gamma ^{\frac{2 c \delta +c}{8 \delta }} \left(8 C_1^2 (2 \delta +1) N^{c/4}+\gamma ^2 (N-6 \delta ) N^{2 \delta }\right)-12 C_1 \delta ^2 \gamma ^{\frac{1}{8} c \left(\frac{1}{\delta }+2\right)+1} N^{\frac{c}{8}+\delta +1}\right.\\ \notag & +256\delta \left(-3 \delta  N \gamma ^{\frac{2 c \delta +c}{8 \delta }} \left(8 C_1^2 (2 \delta +1) N^{c/4}+\gamma ^2 (N-6 \delta ) N^{2 \delta }\right)-12 C_1 \delta ^2 \gamma ^{\frac{1}{8} c \left(\frac{1}{\delta }+2\right)+1} N^{\frac{c}{8}+\delta +1}\right.\\ \notag & \left.\left.+\gamma ^2 \left(4 (N+1) N^2-3 \delta ^2 (N+3)+\delta  (N (7 N+12)-6)\right)\right)\right)\\ \notag & \left.\left.+\gamma ^2 \left(4 (N+1) N^2-3 \delta ^2 (N+3)+\delta  (N (7 N+12)-6)\right)\right)\right) \\ \notag &
\left.\left.\left.+\gamma ^2 \left(-\delta  (3 \delta +2)+4 N^2+7 \delta  N\right)\right)\right)\right)\times \\ \notag &
\left.\left.(\delta +1) (\delta -3 N+2) N^{2 \delta }+\gamma ^2 \left(\delta  (3 \delta +2)-4 N^2-7 \delta  N\right)\right)\right)^{-1}\, ,
\end{align}
while the scalar-to-tensor ratio is given in Eq. (\ref{scalartotensorratiofrphitougianniextra}). As we now demonstrate, the theory at hand is viable for a wide range of the parameters $\gamma$, $\delta$ and $c$, which mainly affect the observational indices. For example, if we choose $c=0.01$, $\delta=-2$, $\gamma=5.9\times 10^3$, $C_1=1$ and $C_2=1$, and for $N=60$ $e$-foldings, the spectral index is found to be equal to $n_s=0.96761$, which is compatible with the Planck data (\ref{planckdata}). For the same choices of values, the scalar-to-tensor ratio reads $r=0.000166667$, which is compatible with both the Planck and the BICEP2/Keck-Array constraints (\ref{scalartotensorbicep2}). As it can be checked, the integration constants $C_1$ and $C_2$ do not crucially affect the observational indices. As we mentioned, the compatibility with the observational data can be achieved for a wide range of parameters, so let us demonstrate this by using some illustrative figures. In Fig. \ref{plot5}, we plot the functional dependence of the spectral index as a function of $c$ (upper left plot) for $N=60$, $\delta=-2$, $\gamma=5.9\times 10^3$, $C_1=1$ and $C_2=1$, as a function of $\delta$ (upper right plot), for $c=0.01$, $\gamma=5.9\times 10^3$, $C_1=1$ and $C_2=1$ and finally, as a function of $\gamma$, for $c=0.01$, $\delta=-2$, $C_1=1$ and $C_2=1$ (lower plot). As in the previous cases, the upper red line and lower black curve correspond to the values $n_s=0.9693$ and $n_s=0.9595$. As it can be seen in all the plots, the compatibility with the observational data can be achieved for a wide range of the free parameters, and in some sense this was expected since the theory has too many free parameters.
The same behavior occurs for the scalar-to-tensor ratio too, and in Fig. \ref{plot6}, we plot the scalar-to-tensor ratio $r$, as a function of the parameter $c$, for $N=60$.
By using the relation for $N(t)$, which can be induced by using Eqs. (\ref{nfoldingsdef}) and (\ref{hubbleratecosmictime}), and since that due to the mimetic constraint, $\phi=t$, we can find the potential $V(\phi)$ and the Lagrange multiplier $\lambda (\phi)$ as a function of $\phi$, but we omit these for brevity. Also, for the choices of the variables we adopted in this case, the cosmological evolution is accelerating, as it can be checked by using Eq. (\ref{hubbleratecosmictime}).

%%%%%%%%%%%%%%%%%%%%%%%%%%%%%%%%%%%%%%%%%%%%%%%%%%%%%%%%%%%%%%%%%%%%%%%%%%%%%%%%%%%%%%%%%%%%%%%%%%%%mimetic 2 31_10_2017

\begin{figure}[h]
\centering
\includegraphics[width=18pc]{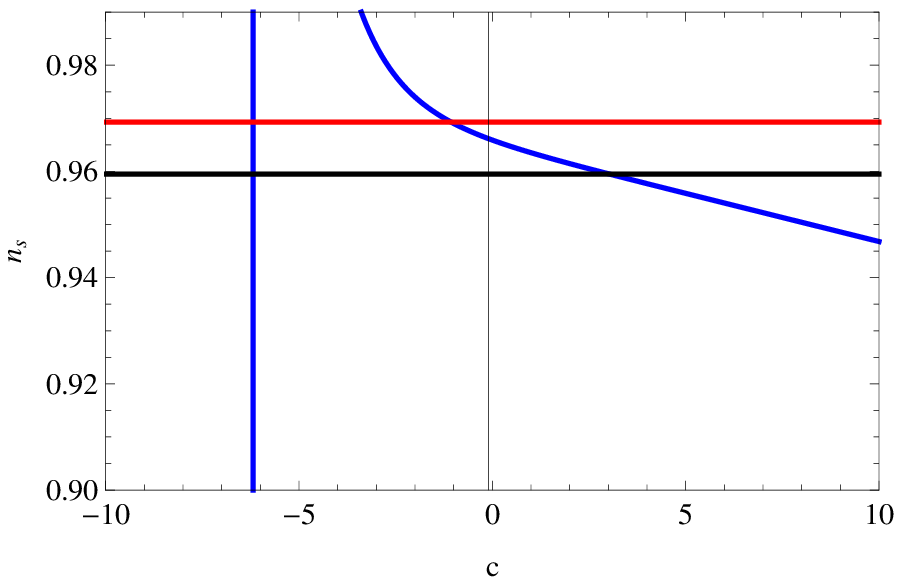}
\includegraphics[width=18pc]{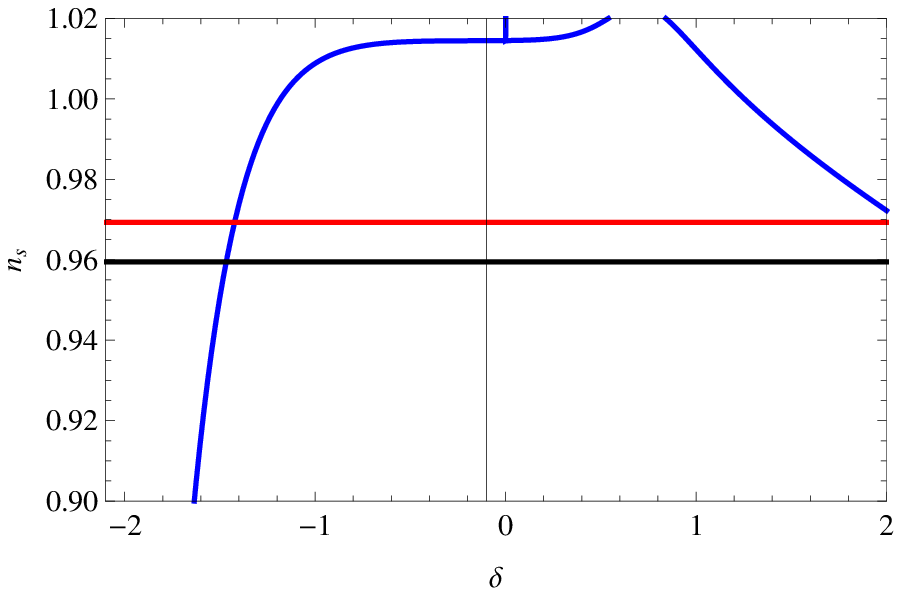}
\includegraphics[width=18pc]{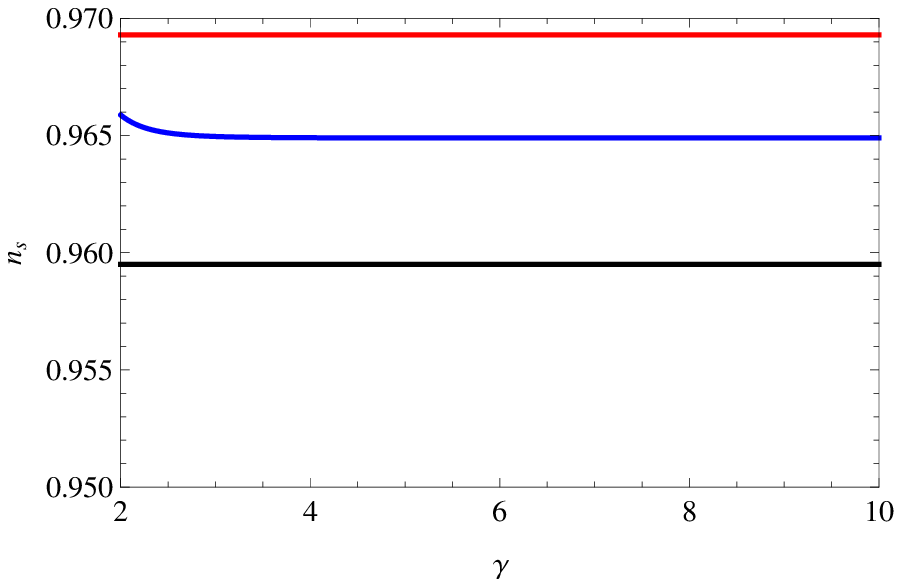}
\includegraphics[width=18pc]{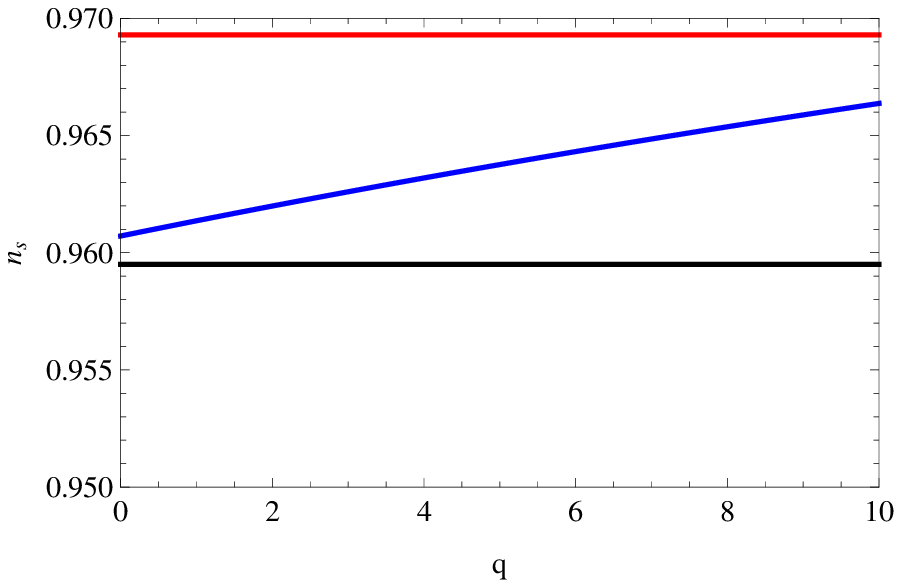}
\caption{The spectral index as a function of $c$ (upper left plot) for $N=60$, $\delta=2.2$, $\gamma=2$, $C_1=1$, $C_2=1$ and $q=9$, as a function of $\delta$ (upper right plot), for $c=0.01$, $\gamma=2$, $q=2$ $C_1=1$, $C_2=1$, as a function of $\gamma$, for $c=0.01$, $\delta=2.2$, $C_1=1$, $C_2=1$ and $q=9$ (lower left plot) and finally as a function of $q$ (lower right), for $c=0.01$, $\delta=2.2$, $\gamma=2$, $C_1=1$ and $C_2=1$.}\label{plot7}
\end{figure}

\begin{figure}[h]
\centering
\includegraphics[width=18pc]{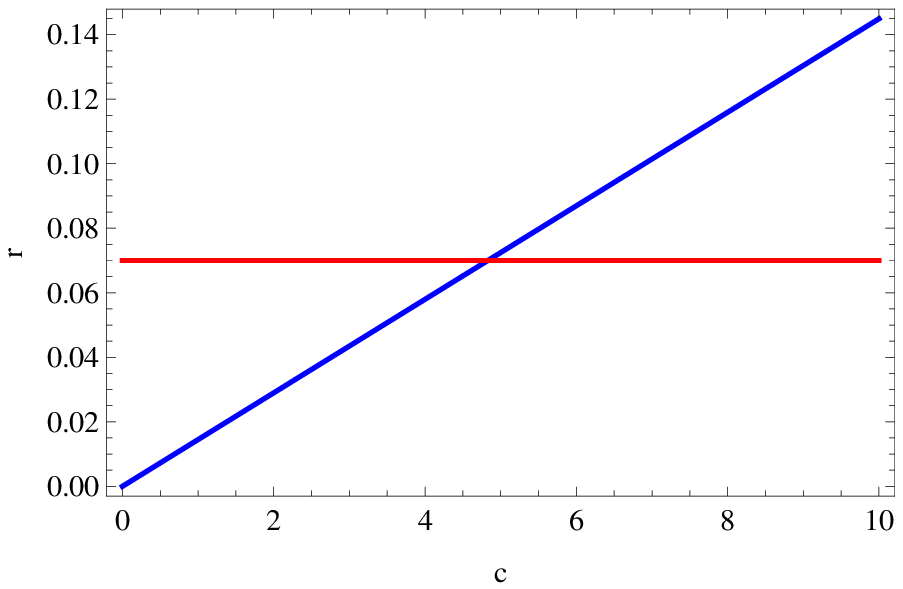}
\includegraphics[width=18pc]{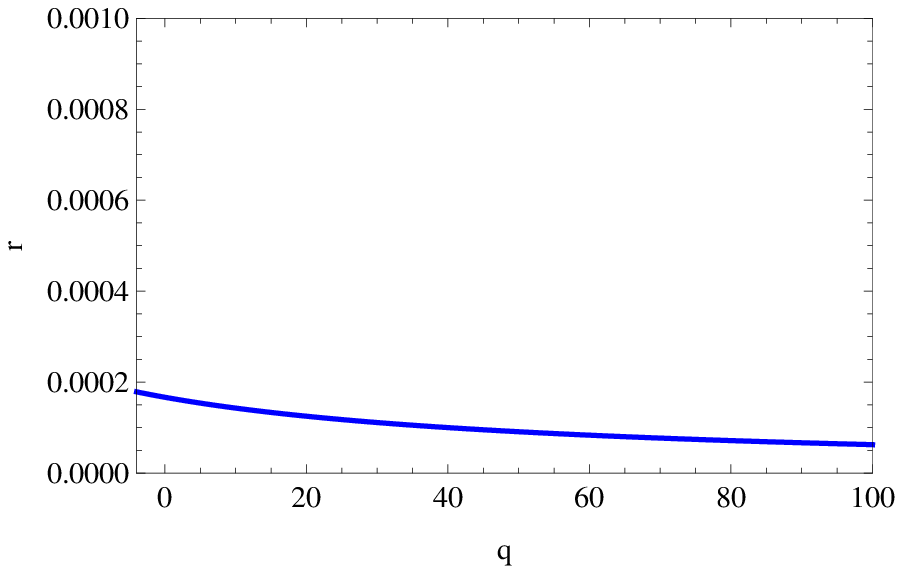}
\caption{The scalar-to-tensor ratio, as a function
of the parameter $c$ (left plot) and as a function of the parameter $q$ (right plot).
}\label{plot8}
\end{figure}

Now let us consider the cosmological evolution appearing in Eq. (\ref{hubbleratemainassumptiontougianni}), and we assume that the scalar-to-tensor ratio has the following form,
\begin{equation}\label{scalartotensorratiofrphitougianniextraglori}
r=\frac{c}{N+q}\, ,
\end{equation}
with $c>0$, so in effect we have $g(N)=\frac{c}{N+q}$ in Eq. (\ref{differentialequationgeneralmimetic}). Thus the differential equation (\ref{differentialequationgeneralmimetic}) becomes in this case,
\begin{equation}\label{differentialequationgeneralmimeticglori}
16 \left(\frac{F'(N)}{2 F(N)}-\frac{H'(N)}{H(N)}\right)=\frac{c}{N+q}\, ,
\end{equation}
and by solving it we obtain the $F(N)$ function, which is,
\begin{equation}\label{fnphirphitougianniextraglori}
F(N)=C_1 (N+q)^{\frac{1}{8} (c-16 \delta )}\, ,
\end{equation}
where $C_1$ is an integration constant. In this case too we assume that the slow-roll condition holds true, so the scalar curvature is approximately $R=12H^2$, so in view of the choice (\ref{hubbleratemainassumptiontougianni}), the scalar curvature is $R\simeq 12 \gamma ^2 (N+q)^{-2 \delta }$, and by inverting we obtain $N(R)=2^{1/\delta } 3^{\frac{1}{2 \delta }} \left(\frac{R}{\gamma ^2}\right)^{-\frac{1}{2 \delta }}-q$, thus the function $F(R)$ is equal to,
\begin{equation}\label{frfinalderivativeglori}
F(R)=C_1 2^{\frac{c}{8 \delta }-2} 3^{\frac{c}{16 \delta }-1} \gamma ^{4 \delta -\frac{c}{4}} R^{1-\frac{c}{16 \delta }}\, ,
\end{equation}
and in effect, the $f(R)$ gravity is,
\begin{equation}\label{caseifrgravityglori}
f(R)=C_2-\frac{R^{2-\frac{c}{16 \delta }} \left(C_1 \delta  2^{\frac{c}{8 \delta }+2} 3^{\frac{c}{16 \delta }-1} \gamma ^{4 \delta -\frac{c}{4}}\right)}{c-32 \delta }\, ,
\end{equation}
where $C_2$ is an integration constant. Accordingly, the function $f(N)$ is equal to,
\begin{equation}\label{fNglori}
f(N)=C_2-\frac{\left(192 C_1 \delta  \gamma ^{2 \left(2-\frac{c}{16 \delta }\right)-\frac{c}{4}+4 \delta }\right) (N+q)^{-2 \delta  \left(2-\frac{c}{16 \delta }\right)}}{c-32 \delta }\, .
\end{equation}
Having $f(N)$ at hand enables us to find the analytic form of the mimetic potential (\ref{breakscalarpotentialefoldings}), which is,
\begin{align}\label{mimeticpotentialofnexplicitform1glori}
& V(N)=-\frac{96 C_1 \delta  \gamma ^{-\frac{2 c \delta +c}{8 \delta }+4 \delta +4} (N+q)^{\frac{c}{8}-4 \delta }}{c-32 \delta }+C_1 \gamma ^{-\frac{c}{4}+4 \delta +4} (N+q)^{-4 \delta -1} (\delta -3 N-3 q) \left(\gamma ^2 (N+q)^{-2 \delta }\right)^{-\frac{c}{16 \delta }}+\frac{C_2}{2}\\ \notag &
+\frac{1}{32} C_1 (c-16 \delta ) \gamma ^{-\frac{c}{4}+4 \delta +2} (N+q)^{-4 \delta -3} \left(\gamma ^2 (N+q)^{-2 \delta }\right)^{-\frac{c}{16 \delta }}\times & \\ \notag &
\left(3 \gamma ^2 \delta ^2+2 \gamma ^2 \delta +8 \gamma ^2 N^2+\delta ^2 (N+q)^{2 \delta }+3 \delta  (N+q)^{2 \delta }+2 (N+q)^{2 \delta }+8 \gamma ^2 q^2\right.\\ \notag &
\left.-2 q \left(5 \gamma ^2 \delta +3 \delta  (N+q)^{2 \delta }+3 (N+q)^{2 \delta }\right)-2 N \left(5 \gamma ^2 \delta +3 \delta  (N+q)^{2 \delta }+3 (N+q)^{2 \delta }-8 \gamma ^2 q\right)\right)\, ,
\end{align}
and accordingly the Lagrange multiplier $\lambda (N)$ is,
\begin{align}\label{lagrangemultiplirerglori}
&  \lambda (N)=\frac{1}{64}C_1\gamma ^{-\frac{c}{4}+4 \delta +2}c-16 \delta (N+q)^{-4 \delta -3}\left(\gamma ^2 (N+q)^{-2 \delta }\right)^{-\frac{c}{16 \delta }}\left((\delta +1) (\delta -3 N+2) (N+q)^{2 \delta }-4 \gamma ^2 q^2\right. \\ \notag & \left.+\gamma ^2 \left(3 \delta ^2-7 \delta  N+2 \delta -4 N^2\right)-q \left(\gamma ^2 (7 \delta +8 N)+3 (\delta +1) (N+q)^{2 \delta }\right)\right)\, .
\end{align}
With the analytic form of $f(N)$, $F(N)$, $V(N)$ and $\lambda (N)$ available, we can proceed to the calculation of the slow-roll indices, and the first three are,
\begin{equation}\label{slowrollanalyticformsglori}
\epsilon_1=\frac{\delta }{N+q},\,\,\,\epsilon_2=0,\,\,\,\epsilon_3=\frac{1}{16} (c-16 \delta ) (N+q)^{\frac{1}{8} (16 \delta -c)+\frac{1}{8} (c-16 \delta )-1}\, ,
\end{equation}
while $\epsilon_4$ is too long to present it here, but it can be calculated accordingly. The spectral index (\ref{obsindices}) in the case at hand can also be calculated, but it has an extended form so we omit for brevity, and the scalar-to-tensor ratio is given in Eq. (\ref{scalartotensorratiofrphitougianniextraglori}). Now let us examine the viability of the model at hand, in which case there are many free parameters. However, the ones which crucially affect the behavior of the observational indices are $c$, $q$, $\gamma$ and $\delta$. The viability comes easily, since there are many free parameters so there is a wide range of available choices. For example, if we choose, $c=0.01$, $\delta=2.2$, $\gamma=2$, $C_1=1$, $C_2=1$ and $q=9$ the spectral index is $n_s=0.965881$, which is within the range of the Planck constraints (\ref{planckdata}), and the corresponding scalar-to-tensor ratio is, $r=0.000144928$, which is in concordance with both the Planck and the BICEP2/Keck-Array data (\ref{scalartotensorbicep2}). In Fig. \ref{plot7}, we present the plots of the spectral index as a function of $c$ (upper left plot) for $N=60$, $\delta=2.2$, $\gamma=2$, $C_1=1$, $C_2=1$ and $q=9$, as a function of $\delta$ (upper right plot), for $c=0.01$, $\gamma=2$, $q=2$ $C_1=1$, $C_2=1$, as a function of $\gamma$, for $c=0.01$, $\delta=2.2$, $C_1=1$, $C_2=1$ and $q=9$ (lower left plot) and finally as a function of $q$ (lower right), for $c=0.01$, $\delta=2.2$, $\gamma=2$, $C_1=1$ and $C_2=1$. As in all the previous cases, the upper red line and lower black curve denote the values $n_s=0.9693$ and $n_s=0.9595$.
Also in Fig. \ref{plot8} we plot the behavior of the scalar-to-tensor ratio as a function of the parameter $c$ (left plot) and as a function of the parameter $q$ (right plot), for $N=60$.
As it can be seen in this case too, the scalar-to-tensor ratio is compatible with both the Planck and BICEP2/Keck-Array data. Finally, the potential and the Lagrange multiplier can be expressed as a function of $\phi=t$, but we omit this task for brevity.

\section{Conclusions}

In this paper we extended the theoretical framework of the inflationary $F(R)$ gravity bottom-up reconstruction technique from the observational indices, to other modified gravities, and particularly to $f(\phi) R$ and mimetic $F(R)$ gravities. After we briefly reviewed the essential features of the $F(R)$ gravity case, we discussed how the inflationary reconstruction technique works in the $f(\phi)R$ case. The method in this case works if the Hubble rate and the scalar-to-tensor ratio are fixed to have a particular form, expressed as functions of the $e$-foldings number $N$. Special attention must be paid in the form of the Hubble rate, since this must yield a cosmology which is accelerating, so it satisfies $\ddot{a}>0$. We provided detailed formulas for all the physical quantities relevant to the theory and also we expressed the slow-roll indices as functions of the $e$-foldings number $N$. Accordingly, we assumed the slow-roll condition to hold true and we calculated in detail the spectral index of the primordial curvature perturbations, which we expressed as function of the $e$-foldings number $N$ too. With regard to the viability of the resulting inflationary theory, the scalar-to-tensor ratio was chosen in such a way so that it is compatible with the current observational constraints coming from the Planck data \cite{Ade:2015lrj} but also from  the BICEP2/Keck-Array data \cite{Array:2015xqh}. As we showed, the spectral index can be compatible with the current observational constraints, if the parameters are appropriately chosen. The viability of the theory occurs for a wide range of the free parameters values, and we illustrated this by using various plots of the observational indices as functions of the free parameters. For the $f(\phi)R$ analysis, we used two accelerating cosmological evolutions, and in both cases the resulting inflationary cosmologies are viable, with the viability being guaranteed for a wide range of the free parameters values. Accordingly, we performed the same analysis for the mimetic $F(R)$ gravity, and we showed how the method functions in this case and also we demonstrated that the resulting inflationary cosmologies are viable. Finally, in the Appendix B we address the $F(R)=R$ mimetic gravity case and we discuss the viability of the model. In this case, as we will demonstrate, only the scalar-to-tensor ratio needs to be specified, and the cosmological evolution is reconstructed from this condition. As we will show, in this case too the viability of the model is guaranteed.

The reconstruction technique from the observational indices we developed in this paper can in principle be extended to other modified gravities, however we need to stress that for the cases studied in this paper, the input for the calculations are only the Hubble rate and the scalar-to-tensor ratio. To be more specific, one needs to fix only the Hubble rate and the scalar-to-tensor ratio to have a desirable form, in order to find which theory can produce a viable inflationary evolution. However, if the same method is applied to more complex theories, such as $f(R,\phi)$ or $F(R,G)$ theories, the Hubble rate and the scalar-to-tensor ratio may not suffice to find all the essential elements of the theory. For example, for the $f(R,\phi)$ theory, it is compelling to specify the functional dependence of $f(R,\phi)$ with respect to the Ricci scalar $R$. A particularly interesting scenario which we did not present in this paper is the $R^2f(\phi)$ case, or more generally the case $f(R,\phi)=F(\phi)g(R)$. Another interesting class of models might be $f(R,\phi)=F(R)+g(\phi)$, which to our opinion is more interesting since the $f(R)$ gravity might be chosen to be a viable theory, such as the Starobinsky model $R+\alpha R^2$ \cite{Starobinsky:1982ee}, or similar.

Moreover, the choice of the Hubble rate is in principle arbitrary, with the only constraint being that the resulting theory should be accelerating, that is, the resulting scale factor as a function of the cosmic time must satisfy $\ddot{a}>0$. Also in this paper we used mainly two types of cosmologies and two forms of the scalar-to-tensor ratio. In principle it is possible to combine the Hubble rates and the forms of the scalar-to-tensor ratio, for example the Hubble rate can be chosen to be $H(N)=\gamma N^{-\delta}$ and the scalar-to-tensor ratio can be chosen as $r=\frac{c}{N+q}$, with $q$ being an arbitrary parameter.

Finally, the analysis we performed can easily be extended to other modified gravities, for which the spectral index of the primordial curvature perturbations and the scalar-to-tensor ratio can be expressed in terms of the slow-roll indices. Also it is possible to perform the same analysis for perfect fluid theories. We hope to address some of the above issues in a future work.

\section*{Acknowledgments}

This work is supported by MINECO (Spain), FIS2016-76363-P and by
CSIC I-LINK1019 Project (S.D.O).

\section*{Appendix A: Observational Indices of the $f(R)$ Gravity Case}

In this appendix we shall prove that the scalar-to-tensor ratio in the case of an $f(R)$ gravity is given by Eq. (\ref{epsilonall}). This result is standard and can be found in many texts \cite{reviews1}. Consider a general $f(R,\phi)$ gravity, in which case the action is,
\begin{equation}
\label{actionfrphigeneral}
\mathcal{S}=\int d ^4x\sqrt{-g}\left
(\frac{f(R,\phi)}{2\kappa^2} -\frac{1}{2}\omega (\phi)
g^{\mu\nu}\partial_{\mu}\phi\partial_{\nu}\phi-V(\phi )\right)\, ,
\end{equation}
where $f(R,\phi)$ is a smooth function of the Ricci scalar and
of the non-canonical scalar field. The slow-roll indices are defined \cite{reviews1},
\begin{equation}
\label{slowrollgenerarlfrphi}
\epsilon_1=-\frac{\dot{H}}{H^2}\, ,\quad
\epsilon_2=\frac{\ddot{\phi}}{H\dot{\phi}}\, , \quad
\epsilon_3=\frac{\dot{f}_R}{2Hf_R}\, ,\quad
\epsilon_4=\frac{\dot{E}}{2HE}\, ,
\end{equation}
with $f_R=\frac{\partial f(R,\phi)}{\partial R}$.
Also  $E$ appearing in
Eq.~(\ref{slowrollgenerarlfrphi}) is defined as follows,
\begin{equation}
\label{epsilonfnction}
E=f_R \omega+\frac{3\dot{f}_R^2}{2\kappa^2\dot{\phi}^2}\, .
\end{equation}
In
addition we define the following function,
$Q_s$,
\begin{equation}
\label{qsfunction}
Q_s=\dot{\phi}^2\frac{E}{f_RH^2(1+\epsilon_3)^2}\, ,
\end{equation}
which is essential for the calculation of the scalar-to-tensor ratio
during the slow-roll era for $f(R)$ gravity. Actually, the observational indices in the case that $\epsilon_i\ll 1$,
$i=1,..,4$, are,
\begin{equation}
\label{observatinalindices1}
n_s\simeq
1-4\epsilon_1-2\epsilon_2+2\epsilon_3-2\epsilon_4\, ,\quad
r=8\kappa^2\frac{Q_s}{f_R}\, .
\end{equation}

In the case of $f(R)$ gravity, $\omega=0$ and $\epsilon_2=0$ so the slow-roll indices become,
\begin{equation}
\label{restofparametersfr}
\epsilon_2=0\, ,\quad
\epsilon_1\simeq
 -\epsilon_3\, ,\quad \epsilon_4\simeq
 -3\epsilon_1+\frac{\dot{\epsilon}_1}{H\epsilon_1}\, .
\end{equation}
Also $Q_s$ becomes,
\begin{equation}\label{qsforfrappendix}
Q_s=\frac{3\dot{f}_R^2}{2\kappa^2f_RH^2(1+\epsilon_3)^2}\, ,
\end{equation}
and from the definition of $\epsilon_3=\frac{\dot{f}_R}{2Hf_R}$, where now $f_R=\frac{\partial f(R)}{\partial R}$, the quantity $Q_s$ reads,
\begin{equation}\label{qsfinalforappendix}
Q_s=\frac{6f_R\epsilon_3^2}{\kappa^2(1+\epsilon_3)^2}\, .
\end{equation}
So by substituting $Q_s$ from Eq. (\ref{qsfinalforappendix}) in Eq. (\ref{observatinalindices1}), the scalar-to-tensor ratio reads eventually,
\begin{equation}\label{rscalafinalappendix}
r=48\frac{\epsilon_3^2}{(1+\epsilon_3)^2}\, .
\end{equation}
Due to the slow-roll condition $\epsilon_3\ll 1$, the scalar-to-tensor ratio reads,
\begin{equation}\label{scalartotensorrationfinalexpression}
r\simeq 48 \epsilon_3^2\, ,
\end{equation}
and due to the fact that $\epsilon_1\simeq -\epsilon_3$ from Eq. (\ref{restofparametersfr}), the scalar-to-tensor ratio finally reads,
\begin{equation}\label{finalfinalr}
r\simeq 48 \epsilon_1^2\, .
\end{equation}
This concludes the proof for the expression of the scalar-to-tensor ratio appearing in Eq. (\ref{epsilonall}). It can be checked that in the case of the Starobinsky $R^2$ model, the expression (\ref{finalfinalr}) yields the standard result $r=12/N^2$, see for example the review \cite{reviews1}. We need to note that the expression (\ref{finalfinalr}) provides the Jordan frame expression of the scalar-to-tensor ratio.

\section*{Appendix B:Bottom-up Reconstruction in the Case $f(R)=R$ }

For completeness, we now consider the case $f(R)=R$ and we investigate how the reconstruction method of the previous sections works in this case. The action in the case at hand is,
\begin{equation}\label{actionmimeticfractionnewadditionforfrr}
S=\int \mathrm{d}x^4\sqrt{-g}\left ( R(g_{\mu \nu}))-V(\phi)+\lambda \left(g^{\mu \nu}\partial_{\mu}\phi\partial_{\nu}\phi
+1\right)\right )\, ,
\end{equation}
which can be transformed to ghost-free $F(R)$ gravity, by using the formalism of Ref. \cite{Nojiri:2017ygt}, in which case the ghost-free action is,
\begin{equation}
\label{FRmim5B}
S_{F(R)}= \frac{1}{2\kappa^2}\int d^4 x \sqrt{-g} \left\{  F(\lambda)
+ \left(\lambda - R \right) \partial_\mu \phi \partial^\mu \phi \right\}
\, ,
\end{equation}
and in principle the observational indices should be the same for both theories. As a first comment, the Hubble rate is restricted if the scalar-to-tensor ratio is fixed to take a specific form, and its exact form is solely determined by the differential equation (\ref{scalartotensoratiogeneralmimetic}), given the functional form of the scalar-to-tensor ratio. Actually, since $f(R)=R$, this implies that $F(R)=1$, so $F'(N)=0$ in Eq. (\ref{scalartotensoratiogeneralmimetic}), so the resulting differential equation reads,
\begin{equation}\label{scalartotensoratiogeneralmimeticnewsection}
r= -\frac{16H'(N)}{H(N)}\, .
\end{equation}
Now let us assume that the scalar-to-tensor ratio has a particular desirable functional form, for example,
\begin{equation}\label{newaddtiition1}
r=\frac{c}{N}\, ,
\end{equation}
and now we investigate the theoretical and phenomenological implications of this choice, for the reconstruction method under study. In this case, the differential equation (\ref{scalartotensoratiogeneralmimeticnewsection}) can be solved and it yields the following Hubble rate,
\begin{equation}\label{hubbleratenewsection1}
H(N)=\gamma N^{-c/16}\, ,
\end{equation}
with $\gamma>0$. The general form of the slow-roll indices in this case is,
\begin{equation}\label{solowrolnewsectionnew}
\epsilon_1=-\frac{H'(N)}{H(N)},\,\,\,\epsilon_2=0,\,\,\,\epsilon_3=0,\,\,\,\epsilon_4=-\frac{\lambda '(N)}{\frac{3}{2} H(N)^2 H'(N)^2-2 \lambda (N)}\, ,
\end{equation}
and the corresponding spectral index is,
\begin{equation}\label{spectralindexnewsection}
n_s=\frac{2 \lambda '(N)}{\frac{3}{2} H(N)^2 H'(N)^2-2 \lambda (N)}+\frac{4 H'(N)}{H(N)}+1\, ,
\end{equation}
while the scalar-to-tensor ratio is given in Eq. (\ref{newaddtiition1}). By using Eq. (\ref{hubbleratenewsection1}), the scalar potential (\ref{scalarpot}) reads,
\begin{equation}\label{scalarpotentialnewsection}
V(N)=3 \gamma ^2 N^{-\frac{c}{8}}+\frac{1}{16} c \gamma ^2 N^{-\frac{c}{8}-1}\, ,
\end{equation}
while the Lagrange multiplier (\ref{lagrange}) is equal to,
\begin{equation}\label{lagrangenewsection1}
\lambda (N)=\frac{1}{32} \gamma ^2 N^{-\frac{c}{8}-1} (c+48 N)\, .
\end{equation}
In effect, and by substituting Eq. (\ref{hubbleratenewsection1}) in Eqs. (\ref{solowrolnewsectionnew}) and (\ref{spectralindexnewsection}), we obtain the analytic form of the slow-roll indices,
\begin{equation}\label{solowrolnewsectionnewanalytic}
\epsilon_1=\frac{c}{16 N},\,\,\,\epsilon_2=0,\,\,\,\epsilon_3=0,\,\,\,\epsilon_4=-\frac{2 c N^{c/8} (c+48 N+8)}{-3 c^2 \gamma ^2+32 c N^{\frac{c}{8}+1}+1536 N^{\frac{c}{8}+2}}\, ,
\end{equation}
and the corresponding spectral index is,
\begin{equation}\label{spectralindexnewsection1}
n_s=\frac{3 c^2 \gamma ^2 (c-4 N)-16 N^{\frac{c}{8}+1} \left(40 c N+(c-8) c-384 N^2\right)}{128 N^{\frac{c}{8}+2} (c+48 N)-12 c^2 \gamma ^2 N}\, .
\end{equation}
As it can be checked, the compatibility with both the Planck and the BICEP2/Keck-Array constraints can be obtained for various values of the free parameters. For example, if we choose $N=60$, $c=2$ and $\gamma=1221.59$, the spectral index becomes equal to $n_s=0.966$, while the scalar-to-tensor ratio reads $r=0.0333333$. We can easily find the scalar potential and the Lagrange multiplier as a function of $\phi$, by solving the differential equation $N(t)=\dot{H}(N(t))$, with the Hubble rate $H(N)$ appearing in Eq. (\ref{hubbleratenewsection1}), so the resulting $N(t)$ is,
\begin{equation}\label{ntneswesection1}
N(t)=16^{-\frac{16}{c+16}} ((c+16) (C_1+\gamma  t))^{\frac{16}{c+16}}\, ,
\end{equation}
with $C_1$ being an arbitrary integration constant. By inverting (\ref{ntneswesection1}) and using Eq. (\ref{scalarpotentialnewsection}), the scalar potential is,
\begin{equation}\label{sclarpotentialexplicitform1}
V(\phi)=2^{-\frac{64}{c+16}-3} \gamma ^2 \left(2^{-\frac{64}{c+16}} ((c+16) (\gamma  \phi +C_1))^{\frac{16}{c+16}}\right)^{-\frac{c}{8}-1} \left(48 ((c+16) (\gamma  \phi +C_1))^{\frac{16}{c+16}}+2^{\frac{64}{c+16}} c\right)\, ,
\end{equation}
and also by using (\ref{lagrangenewsection1}), the Lagrange multiplier reads,
\begin{equation}\label{lagrangeexplicitform1}
\lambda (\phi)=\frac{1}{32} \gamma ^2 \left(2^{-\frac{64}{c+16}} ((c+16) (\gamma  \phi +C_1))^{\frac{16}{c+16}}\right)^{-\frac{c}{8}-1} \left(3\ 16^{\frac{c}{c+16}} ((c+16) (\gamma  \phi +C_1))^{\frac{16}{c+16}}+c\right)\, .
\end{equation}

Let us now investigate various alternative forms of the scalar-to-tensor ratio, so let us start with the choice,
\begin{equation}\label{newaddtiition1newaddition1}
r=\frac{c}{N^2}\, .
\end{equation}
In this case, the differential equation (\ref{scalartotensoratiogeneralmimeticnewsection}) yields the following Hubble rate,
\begin{equation}\label{hubbleratenewsection1newaddition1}
H(N)=\gamma  e^{\frac{c}{16 N}}\, ,
\end{equation}
with $\gamma>0$. The corresponding scalar potential (\ref{scalarpot}) reads,
\begin{equation}\label{scalarpotentialnewsectionnewaddition1}
V(N)=\frac{c \gamma ^2 e^{\frac{c}{8 N}}}{16 N^2}+3 \gamma ^2 e^{\frac{c}{8 N}}\, ,
\end{equation}
while the Lagrange multiplier (\ref{lagrange}) is in this case,
\begin{equation}\label{lagrangenewsectioncaseii}
\lambda (N)=\frac{\gamma ^2 e^{\frac{c}{8 N}} \left(c+48 N^2\right)}{32 N^2}\, .
\end{equation}
The corresponding slow-roll indices are,
\begin{equation}\label{solowrolnewsectionnewanalyticnewaddition1}
\epsilon_1=\frac{c}{16 N^2},\,\,\,\epsilon_2=0,\,\,\,\epsilon_3=0,\,\,\,\epsilon_4=\frac{2 c (c+16 N (3 N+1))}{3 c^2 \gamma ^2 e^{\frac{c}{8 N}}-32 N^2 \left(c+48 N^2\right)}\, ,
\end{equation}
and the corresponding spectral index is,
\begin{equation}\label{spectralindexnewsection1newaddition1}
n_s=1-\frac{c}{4 N^2}-\frac{4 c (c+16 N (3 N+1))}{3 c^2 \gamma ^2 e^{\frac{c}{8 N}}-32 c N^2-1536 N^4}\, .
\end{equation}
As it can be checked, in this case too, the observational indices can be compatible with the observational data for various choices of the free parameters. For example, by choosing $N=60$, $c=1$ and $\gamma=81416$, then we obtain $n_s=0.966$, and the scalar-to-tensor ratio is $r=0.000277778$ and thus the observational indices are compatible with both the Planck and the BICEP2/Keck-Array data.

Finally, let us assume that,
\begin{equation}\label{newaddtiition1newaddition12}
r=\frac{c}{N+q}\, ,
\end{equation}
in which case, the Hubble rate is,
\begin{equation}\label{hubbleratenewsection1newaddition12}
H(N)=\gamma  (N+q)^{-\frac{c}{16}}\, ,
\end{equation}
with $\gamma>0$. The corresponding scalar potential (\ref{scalarpot}) reads,
\begin{equation}\label{scalarpotentialnewsectionnewaddition12}
V(N)=3 \gamma ^2 (N+q)^{-\frac{c}{8}}+\frac{1}{16} c \gamma ^2 (N+q)^{-\frac{c}{8}-1}\, ,
\end{equation}
and the Lagrange multiplier (\ref{lagrange}) is,
\begin{equation}\label{lagrangemultiplielastcase}
\lambda (N)=\frac{1}{32} \gamma ^2 (N+q)^{-\frac{c}{8}-1} (c+48 (N+q))
\end{equation}
Hence the slow-roll indices are,
\begin{equation}\label{solowrolnewsectionnewanalyticnewaddition12}
\epsilon_1=\frac{c}{16 (N+q)},\,\,\,\epsilon_2=0,\,\,\,\epsilon_3=0,\,\,\,\epsilon_4=-\frac{2 c (N+q)^{c/8} (c+48 N+48 q+8)}{32 (N+q)^{\frac{c}{8}+1} (c+48 (N+q))-3 c^2 \gamma ^2}\, ,
\end{equation}
and the corresponding spectral index is,
\begin{align}\label{spectralindexnewsection1newaddition12}
& n_s=\frac{6144 (N+q)^{\frac{c}{8}+3}-128 c \left(5 N^2+N (10 q-1)+q (5 q-1)\right) (N+q)^{c/8}}{4 (N+q) \left(-3 c^2 \gamma ^2+32 c (N+q)^{\frac{c}{8}+1}+1536 (N+q)^{\frac{c}{8}+2}\right)}\\ \notag &
+\frac{3 c^3 \gamma ^2-4 c^2 (N+q) \left(3 \gamma ^2+4 (N+q)^{c/8}\right)}{4 (N+q) \left(-3 c^2 \gamma ^2+32 c (N+q)^{\frac{c}{8}+1}+1536 (N+q)^{\frac{c}{8}+2}\right)}
\, .
\end{align}
By making the choice $N=60$, $c=1$, $\gamma=1845.24$ and $q=1$, we get $n_s=0.966$ and $r=0.0163934$, so in this case too, the results are compatible with both the Planck and the BICEP2/Keck-Array data. Finally, the potential $V(\phi)$ and the Lagrange multiplier $\lambda (\phi)$ can be found by finding $N(t)$ in this case, for the Hubble rate (\ref{hubbleratenewsection1newaddition12}), which is,
\begin{equation}\label{ntfinalquotelast}
N(t)=16^{-\frac{16}{c+16}} ((c+16) (C_1+\gamma  t))^{\frac{16}{c+16}}\, ,
\end{equation}
so the resulting scalar potential and the Lagrange multiplier are identical to the ones given in Eqs. (\ref{sclarpotentialexplicitform1}) and (\ref{lagrangeexplicitform1}). Therefore, the cosmological evolutions (\ref{hubbleratenewsection1}) and (\ref{hubbleratenewsection1newaddition12}) are degenerate at the Lagrangian level. Hence the $f(R)=R$ case yields interesting results. We need to note that we used a different approach in comparison with Ref. \cite{Chamseddine:2014vna}, where the authors also found interesting inflationary solutions, but with specifying the scalar potential. In our approach we started from the scalar-to-tensor ratio by specifying it's functional form.

\section*{Appendix C: The Calculation of the Scalar-to-tensor Ratio for the Mimetic $f(R)$ Gravity}

Here we will prove that the scalar-to-tensor ratio in the case of a mimetic $f(R)$ gravity is equal to,
\begin{equation}\label{rscalarappendixmimetic}
r\simeq 16(\epsilon_1+\epsilon_2)\, ,
\end{equation}
when the slow-roll approximation holds true. Recall that the slow-roll conditions are,
\begin{equation}
\label{slowrollerarealtionsrevision}
\dot{H}\ll H^2\, ,\quad \ddot{H}\ll H\dot{H}\, ,
\end{equation}
The mimetic $f(R)$ gravity with the constraint (\ref{enm4}), which yields $\phi=t$, can be considered as a generalized $f(R,\phi)$ scalar-tensor theory, with action given in Eq. (\ref{MF1sc}). For convenience we quote here the action, which is,
\be
\label{MF1screvision}
S = \int \sqrt{-g} \left\{ \frac{f(R)}{2\kappa^2} + \lambda (\phi) \partial_\mu \phi \partial^\mu \phi
+ \lambda(\phi ) - V(\phi) \right\} \, ,
\ee
hence the kinetic term is $\omega (t)=-2 \lambda (t )$, and the scalar potential $\mathcal{V}(t)$ is $\mathcal{V}(t)=\lambda (t)-V(t)$. By varying with respect to the metric and with respect to the scalar field, we get the following equations of motion,
\begin{align}
\label{eqnsofmotionfrphinonslowrollrevision}
& \frac{3F(R,\phi)H^2}{\kappa^2}=\frac{\omega
(\phi)\dot{\phi}^2}{2}+V(\phi)+\frac{1}{\kappa^2}\left(
RF(R,\phi)-f(R,\phi)\right)-3H\dot{F}(R,\phi) \, , \nn
& -\frac{2F(R,\phi)}{\kappa^2}\dot{H}=\omega
(\phi)\dot{\phi}^2+\ddot{F}(R,\phi)-H\dot{F}(R,\phi) \, ,\nn
& \ddot{\phi}+3H\dot{\phi}
+\frac{1}{2\omega (\phi)}\left( \dot{\omega
(\phi)}\dot{\phi}-\frac{1}{\kappa^2}\frac{\partial
f(R,\phi)}{\partial \phi}+2\frac{\partial V}{\partial
\phi}\right)=0\, ,
\end{align}
where $F=\frac{\partial f}{\partial R}$. The second equation, in the slow-roll approximation, and due to the fact that $\omega=-2\lambda$ can be approximated as follows,
\begin{equation}\label{secondeqn}
-2F(R,\phi)\dot{H}\simeq -2\lambda
(\phi)\dot{\phi}^2-H\dot{F}(R,\phi)\, ,
\end{equation}
where have set $\kappa^2=1$ for convenience. The first and third slow-roll indices are,
\begin{equation}
\label{slowrollindicesrevision}
\epsilon_1=-\frac{\dot{H}}{H^2}\, , \epsilon_3=\frac{\dot{F}(R,\phi)}{2HF(R,\phi)}\, ,
\end{equation}
where $E(R,\phi)$ is equal to,
\begin{equation}
\label{epsilonfunctionrevision}
E(R,\phi)=F(R,\phi)\omega (\phi)+\frac{3\dot{F}(R,\phi)^2}{2\dot{\phi}^2}\, ,
\end{equation}
which due to the mimetic constraint identification $\phi=t$, becomes,
\begin{equation}
\label{epsilonfunctionrevision1}
E(R,\phi)=-2F(R,\phi)\lambda (\phi)+\frac{3\dot{F}(R,\phi)^2}{2}\, .
\end{equation}
This can be further expressed as follows,
\begin{equation}
\label{epsilonfunctionrevision2}
E(R,\phi)=-2F(R,\phi)\lambda (\phi)+\frac{3F_{RR}\dot{R}^2}{2}\, ,
\end{equation}
where $F_{RR}=\frac{\partial^2 f}{\partial R^2}$. Due to the slow-roll approximation, the dominant behavior of $\dot{R}^2$ is $\dot{R}^2\sim \dot{H}^2H^2$, so this term can be omitted, so we eventually have,
\begin{equation}
\label{epsilonfunctionrevision3}
E(R,\phi)\simeq -2F(R,\phi)\lambda (\phi)\, .
\end{equation}
The scalar-to-tensor ratio is generally given by,
\begin{equation}\label{scalarmimeticratio}
r=\frac{8Q_s}{F}\, ,
\end{equation}
where $Q_s$ is,
\begin{equation}\label{qsforapprendixmimetic}
Q_s=\frac{E}{FH^2}\, .
\end{equation}
Substituting Eq. (\ref{secondeqn}) in Eq. (\ref{epsilonfunctionrevision3}), we obtain,
\begin{equation}\label{qsexpression}
Q_s\simeq -\frac{2F\dot{H}}{H^2}+\frac{\dot{F}}{H}\, ,
\end{equation}
so by substituting (\ref{qsexpression}) in the scalar-to-tensor ratio (\ref{scalarmimeticratio}), we obtain,
\begin{equation}\label{prefinalexpression}
r\simeq 8\left( -\frac{2\dot{H}}{H^2}+\frac{\dot{F}}{HF}\right)\, ,
\end{equation}
and by using the expressions for the slow-roll indices $\epsilon_1$ and $\epsilon_3$ from Eq. (\ref{slowrollindicesrevision}), we obtain that $r\simeq 16 (\epsilon_1+\epsilon_3)$, which is the exact expression appearing in Eq. (\ref{rscalarappendixmimetic}).

\end{document}